\documentclass[journal]{IEEEtran}

\usepackage{xcolor}
\usepackage{xpatch}
\usepackage{booktabs,bbding}
\usepackage{threeparttable}
\usepackage{amsmath,graphicx,cite,amssymb}
\usepackage{amsfonts,multirow,bm,array,setspace,stfloats}

\usepackage{graphicx,float,cite,amssymb}
\usepackage{graphics}\DeclareGraphicsExtensions{.pdf,.jpeg,.png,.jpg}   
\usepackage{multirow,bm,bbm,array,setspace,dsfont}
\usepackage{textcomp}
\usepackage{psfrag}
\usepackage{color}
\usepackage{pstricks,enumerate}
\usepackage{bbm,subfigure}
\usepackage[lined,boxed,commentsnumbered]{algorithm2e}
\usepackage{lipsum,booktabs}
\usepackage{makecell}

\usepackage{psfrag}
\usepackage{epsfig}

\makeatletter
\newcommand{\distas}[1]{\mathbin{\overset{#1}{\kern\z@\sim}}}%
\newsavebox{\mybox}\newsavebox{\mysim}
\newcommand{\distras}[1]{%
  \savebox{\mybox}{\hbox{\kern1pt$\scriptstyle#1$\kern1pt}}%
  \savebox{\mysim}{\hbox{$\sim$}}%
  \mathbin{\overset{#1}{\kern\z@\resizebox{\wd\mybox}{\ht\mysim}{$\sim$}}}%
}

\ExplSyntaxOn
\cs_new:Npn \bibColoredItems #1#2
  {
    \clist_map_inline:nn {#2} { \cs_new:cpn {bib@colored@##1} {#1} } 
  }
\ExplSyntaxOff

\newcommand\bib@setcolor[1]{%
  \ifcsname bib@colored@#1\endcsname
    \expandafter\color\expandafter{\csname bib@colored@#1\endcsname}
  \else
    \normalcolor
  \fi
}

\xpatchcmd\@bibitem
  {\item}
  {\bib@setcolor{#1}\item}
  {}{\PatchFailed}

\xpatchcmd\@lbibitem
  {\item}
  {\bib@setcolor{#2}\item}
 {}{\PatchFailed}
\makeatother
\ifCLASSINFOpdf
\else
\fi
\newtheorem{example}{Example}

\newcommand{\bA}{\bm A}

\newcommand{\bB}{\bm B}
\newcommand{\bd}{\bm d}

\newcommand{\bg}{\bm g}

\newcommand{\bx}{\bm{x}}
\newcommand{\bS}{\bm{S}}
\newcommand{\bs}{\bm{s}}
\newcommand{\bI}{\bm{I}}

\newcommand{\bv}{\bm v}
\newcommand{\bV}{\bm V}

\newcommand{\bY}{\bm{Y}}
\newcommand{\by}{\bm{y}}

\newcommand{\bP}{\bm{P}}

\newcommand{\bH}{\bm{H}}
\newcommand{\bh}{\bm{h}}
\newcommand{\bW}{\bm{W}}
\newcommand{\bD}{\bm{D}}
\newcommand{\bz}{\bm{z}}

\newcommand{\bw}{\bm{w}}
\newcommand{\bZ}{\bm{Z}}

\newcommand{\hh}{\mathrm{H}}

\newcommand{\notleftright}{\mathrel{\ooalign{$\Longleftrightarrow$\cr\hidewidth$/$\hidewidth}}}

\begin{document}

\title{Quadratic Transform for Fractional Programming in Signal Processing and Machine Learning}
\author{
	\IEEEauthorblockN{
	Kaiming Shen, \IEEEmembership{Senior Member,~IEEE} and Wei Yu, \IEEEmembership{Fellow,~IEEE}
}\thanks{
Manuscript submitted to IEEE Signal Processing Magazine on August 3, 2024, revised on December 21, 2024 and March 13, 2025, and accepted on March 24, 2025.

Kaiming Shen is with the School of Science and Engineering, The Chinese University of Hong Kong (Shenzhen), Shenzhen 518172, China (e-mail: shenkaiming@cuhk.edu.cn).

Wei Yu is with the Edward S. Rogers Sr.
Department of Electrical and Computer Engineering, University of Toronto,
Toronto, ON M5S 3G4, Canada (e-mail: weiyu@ece.utoronto.ca).
} % <-this % stops a space
}

%Telephone: (800) 555--1212, Fax: (888) 555--1212}


\maketitle

%\vspace{-5em}
\begin{abstract}
Fractional programming (FP) is a branch of mathematical optimization that deals with
the optimization of ratios. It is an invaluable tool for signal processing and machine learning, 
because many key metrics in these fields are fractionally structured, e.g., the signal-to-interference-plus-noise ratio (SINR) in wireless communications, the Cram\'{e}r-Rao bound (CRB) in radar sensing, the normalized cut in graph clustering, and the margin in support vector machine (SVM). This article provides a comprehensive review of both the theory and applications of a recently developed FP technique known as the \emph{quadratic transform}, which can be applied to a wide variety of FP problems, including both the minimization and the maximization of the sum of functions of ratios as well as matrix ratio problems.
\end{abstract}

\section{Introduction}\label{sec:overview}

\IEEEPARstart{F}{ractional} 
programming (FP) refers to optimization problems involving ratios. Consider for example the maximization of the sum of multiple ratios 
over a vector $x \in \mathbb{R}^d$:
\begin{equation}
\underset{x \in \mathcal{X}}{\text{maximize}} \quad f(x) := \sum^n_{i=1} \frac{A_i(x)}{B_i(x)},
\label{eq:intro}
\end{equation}
where $A_i(x) \ge 0$ and $B_i(x) > 0$ are functions of $x$, and $\mathcal X$ is a constraint set. While one could simply 
treat $f(x)$ as a generic function and use, e.g., gradient-based method, to 
maximize the objective, better approaches are possible if we exploit the 
fractional structure of $f(x)$. Generic approaches for optimizing fractions do not
always work well, because $B_i(x)$ can have strong
influence on the overall objective value when $B_i(x)$ is close to zero. Consequently, 
it is not always easy to find the appropriate step size when optimizing the objective
with the numerators $A_i(x)$ and the denominators $B_i(x)$ coupled in 
the fractional form.  An important idea in the study of FP problems is to decouple 
the numerators and the denominators and to put them inside different terms in a summation.  Specifically, as we wish to maximize $A_i(x)$
while minimizing $B_i(x)$ at the same time, it would be sensible to maximize
the sum of an \emph{increasing} function of $A_i(x)$ and a \emph{decreasing}
function of $B_i(x)$.  The choices of these \emph{transformed} objectives would
affect the general applicability and performance of the methods. Many of the
classical and modern FP techniques hinge on the judicious
choices of these functions in the transform. Classical
methods in FP, such as the Charnes-Cooper method \cite{charnes_1962,schaible_transform}, Dinkelbach's method \cite{dinkelbach_transform}, as well as the method in \cite{Jibetean} that transforms the optimization of a rational function into a hierarchy of semidefinite programming (SDP) relaxations, all have such transforms embedded in their core ideas.

This article provides a comprehensive review of a recently developed technique called the quadratic
transform \cite{shen2018fractional} to solve FP problems. The main idea of quadratic transform is to introduce
a set of auxiliary variables $y_i$ and to transform the maximization of the sum of ratios over $x$ as in \eqref{eq:intro} to a joint maximization over $x$ and $y_i$ as follows:
%sequentially (over $y_i$) as follows:
\begin{equation}
  \label{prob 2}
\underset{x \in \mathcal{X},\, y_1,\cdots,y_n}{\text{maximize}}\quad \sum^n_{i=1} \Big(2y_i\sqrt{A_i(x)}-y^2_iB_i(x)\Big),
\end{equation}
which can then be solved numerically by iterating between $x$ and $y_i$'s.
The equivalence between \eqref{eq:intro} and \eqref{prob 2} can be established by explicitly optimizing over $y_i$ for fixed $x$ and substituting the optimal $y_i$ back into the objective function of \eqref{prob 2}. The key advantage of the quadratic transform is that it allows the
decoupling of numerator and denominator for more than one ratio simultaneously in multi-ratio FP such as \eqref{eq:intro}, whereas the
classical FP methods such as the Charnes-Cooper and Dinkelbach's
methods can decouple only a single ratio, as we shortly explain.

\begin{figure}[t]
\psfrag{x}[c][c][0.7]{$uiop$}
\centering
\includegraphics[width=0.8\linewidth]{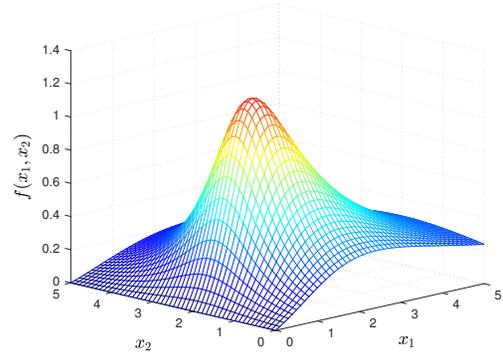}
\caption{An example of a single-ratio objective function $f(x_1,x_2)=\frac{x_1}{(x_1-1)^2+(x_2-1)^2+1}$ that satisfies the concave-convex condition and thus is quasi-concave, for which a local optimum is also the global optimum.}
\label{fig:concave_convex}
\end{figure}

The study on FP was initiated by John von Neumann in 1937 in his seminal
work on economic equilibrium. It has since been considered extensively in
broad areas including economics, management science, information theory,
optics, graph theory, and computer science \cite{IM_fp}. Many early works
focus on the single-ratio problem under the assumption that the numerator is a concave
function and the denominator is a convex function (known as the
\emph{concave-convex condition}), so that the optimization objective is
quasi-concave, as shown by an example in Fig.~\ref{fig:concave_convex}. 
The quasi-concave structure of the single-ratio FP allows the development
of classic methods, namely the Charnes-Cooper method
\cite{charnes_1962,schaible_transform} and Dinkelbach's method
\cite{dinkelbach_transform} for achieving the global optimum of the
single-ratio problem efficiently. These two classic methods have long been
recognized as standard tools for FP, but neither of them can be
extended to multi-ratio problems (except for the max-min-ratios case
\cite{crouzeix1985algorithm}). In fact, because even the basic
sum-of-ratios problem can be shown to be NP-complete \cite{freund}, many 
of the past studies for the multi-ratio FP have focused on the global optimization approaches such as branch-and-bound, which have an exponential worst-case complexity. Consequently, early
applications of FP in communications and signal processing \cite{jorswieck}
are often restricted to either problems of small size, or problems that contain only a single ratio, e.g., the
maximization of energy efficiency in a wireless network.

\begingroup
\begin{table*}[t]
\renewcommand{\arraystretch}{2.0}
\centering
\caption{\small Different types of FP problems and their applications}
\begin{tabular}{|l|l|l|}
\firsthline
FP Problems & Quadratic Transform Methods & Examples of Applications \\
\hline
\hline
\makecell*[l]{single-ratio:\\$\max_x A(x)/B(x)$} & quadratic transform \cite{shen2018fractional} & link-level energy efficiency \\
\hline
\makecell*[l]{max-min-ratios:\\$\max_x\min_iA_i(x)/B_i(x)$}  & quadratic transform \cite{shen2018fractional} & \makecell*[l]{support vector machine (SVM),\\ system-level energy efficiency}\\
\hline
\makecell*[l]{sum-of-ratios max:\\$\max_x\sum_iA_i(x)/B_i(x)$} & quadratic transform \cite{shen2018fractional} &\makecell*[l]{power control in cellular networks}\\
\hline
\makecell*[l]{sum-of-ratios min:\\$\min_x\sum_iA_i(x)/B_i(x)$}  & inverse quadratic transform \cite{Yannan_TSP24} &age-of-information (AoI) minimization\\
\hline
\makecell*[l]{sum-of-functions-of-ratio:\\$\max_x\sum_if_i(A_i(x)/B_i(x))$}  &unified quadratic transform \cite{Yannan_TSP24} & \makecell*[l]{secure transmission}\\
\hline
\makecell*[l]{sum-of-log-ratios:\\$\max_x\sum_i\log(1+A_i(x)/B_i(x))$} &logarithmic quadratic transform \cite{shen2018fractional2} & \makecell*[l]{joint power control and user scheduling}\\
\hline
\makecell*[l]{matrix-ratio:\\extend ${A_i(x)}/{B_i(x)}$ to matrix\\ $\sqrt{\bA_i(x)}^\hh\bB^{-1}_i(x)\sqrt{\bA_i(x)}$} & \makecell*[l]{matrix quadratic transform \cite{shen2019optimization},\\nonhomogeneous quadratic transform \cite{ZP_MM+},\\extrapolated quadratic transform \cite{Shen_JSAC24}} & \makecell*[l]{ unsupervised data clustering,\\ multi-antenna beamforming,\\ pilot design for channel estimation}\\
\lasthline
\end{tabular}
\label{tab:application}
\end{table*}
\endgroup

A basic tenet in optimization is that for an algorithm to be efficient, it
should take advantage of the problem structure; generic optimization methods
(e.g., gradient descent), which can be applied independently of the specific
problem structure, would not be as efficient. By focusing attention on the
fractional structure---which is a common characteristic of many 
signal processing and machine learning problems and is often the most important part of their main features, the quadratic transform is able 
to capture the problem-specific features while being generically applicable
at the same time. Intuitively, the quadratic transform leverages the
fractional structure of the problem while leaving other parts of the
problem generic, in order to reach a desirable tradeoff between
universality and exploitation of problem-specific structures.

This article starts by treating the single-ratio problem, 
the max-min-ratios problem, and the sum-of-ratios problem,  
then progresses toward a wider range of FP problems, including the sum-of-functions-of-ratio, the sum-of-log-ratios, and the matrix-ratio problems, as summarized 
in Table \ref{tab:application} along with their application areas.
This feature article restricts attention to optimization problems with a single
objective, although multi-objective FP \cite{multiobj_FP} and bilevel FP
\cite{bilevel_FP} have also been considered in the literature. The focus of
this article is on the quadratic transform and its relatives; we leave interested
readers to consult additional literature \cite{jaberipour2010solving,evolutionary} for other heuristic approaches for solving FP, e.g., the harmony search and the evolutionary algorithm.
Table \ref{tab:methods} contains a summary of the features of the quadratic
transform as compared to other commonly used optimization methods for FP problems under the concave-convex or the convex-concave condition.

In terms of the application area, the ubiquity of SINR is a
main motivation for treating many communications and signal processing
problems from an FP perspective. Unlike metrics such as the overall energy efficiency, SINRs are often
measured at more than one entity (e.g., at multiple receivers or for different
signals) in a communication system, so the
corresponding problem is typically a multi-ratio FP. Recently, there have been a
flurry of efforts in applying the quadratic transform to different research frontiers in wireless communications, e.g., cell-free massive multiple-input multiple-output (MIMO) system
\cite{Bjornson_cellfree21}, satellite network \cite{Gu_TWC22_satellite},
and intelligent reflecting surface (IRS) or reconfigurable intelligent
surface (RIS) \cite{larsson_RIS} systems. Other similar metrics, e.g., the
signal-to-leakage-plus-noise ratio (SLNR) \cite{Zhang_TSP20_mmWave}, can 
also be handled by the quadratic transform. Aside from the communications
problems, the quadratic transform has been applied to the wireless sensing
problems as well. In particular, there has been extensive recent research interest
in using the quadratic transform to maximize the communication SINR and the
radar SINR jointly for integrated sensing and communications (ISAC)
applications, e.g., in \cite{masouros_isac_twc22}. Other commonly used metrics
for radar signal processing include the CRB
\cite{Yannan_TSP24} and the ambiguity function sidelobe level ratio
\cite{Yang_TVT20_ISAC}, both of which are fractionally structured and hence
amenable to the quadratic transform based optimization. In the area of
image processing, \cite{Palomar_TSP19_shaping} uses Dinkelbach's method to
handle the spectral level ratio for medical imaging,
\cite{Lei_TCI24_imaging} uses Dinkelbach's method to maximize the ratio of
data consistency to the regulation term for electrical capacitance
tomography, and \cite{Beck_SIAM06_QCFQ} proposes a novel FP approach to the regularized total least squares (RTLS) problem for image deblurring.
Latency is another key metric with a fractional structure. The quadratic
transform has been adopted to reduce latency in cloud radio access networks
(C-RAN) \cite{Park_edge_computing_21} and in federated edge learning
systems \cite{guanding_FL}. 

Machine learning is a field of ever increasing importance where FP
has also found ample applications.  SVM is a
fundamental and popular supervised classifier. It aims to maximize the
distance between the boundary and the nearest data points, named
\emph{margin}, in order to minimize the misclassification error. 
The margin maximization problem is nonconvex. A typical solution technique
is to reformulate the problem into a convex form, then apply the
Lagrangian duality theory. But since the margin has a fractional 
structure, the SVM problem can be directly solved by Dinkelbach's 
method. A potential advantage of using the FP approach for SVM is that 
it may be able to deal with multi-class SVM problems (which are multi-ratio FP \cite{SVM_FP}), for which the classical method cannot 
be easily applied.

\begingroup
\begin{table*}[t]
%\footnotesize
\renewcommand{\arraystretch}{1.8}
\centering
\caption{\small Comparison of the different methods for FP under the concave-convex or the convex-concave condition}
\begin{tabular}{|c|c|c|c|c|}
\firsthline
  & \makecell*[c]{Charnes-Cooper\\ \cite{charnes_1962,schaible_transform}}& \makecell*[c]{Dinkelbach\\ \cite{dinkelbach_transform,crouzeix1985algorithm}}& \makecell*[c]{Quadratic Transform\\ \cite{shen2018fractional,Yannan_TSP24,shen2018fractional2,shen2019optimization,ZP_MM+,Shen_JSAC24}} & \makecell*[c]{AM-GM\\Inequality \cite{ZhaoJun_JSAC}} \\
\hline
\hline
single-ratio & global optimum & global optimum & global optimum & global optimum \\
\hline
max-min-ratios & N/A & global optimum & global optimum & global optimum \\
\hline
sum-of-ratios max & N/A & N/A & stationary point & N/A \\
\hline
sum-of-ratios min & N/A & N/A & stationary point & stationary point\\
\hline
sum-of-log-ratios & N/A & N/A & stationary point & N/A \\
\hline
matrix-ratio & N/A & N/A & stationary point & N/A \\
\hline
\hline
convergence rate & \makecell*[c]{iterations\\ not required} & superlinear & \makecell*[c]{slower than\\ Dinkelbach} & \makecell*[c]{slower than\\ Dinkelbach} \\
\hline
\lasthline
\end{tabular}
%   \begin{tablenotes}
%   \centering
%      \small
%      \item Note: \XSolidBrush \, indicates that the method is not applicable to the problem.
%    \end{tablenotes}
\label{tab:methods}
\end{table*}
\endgroup

Aside from supervised classification, the unsupervised clustering 
problem is also closely related to FP, because the commonly used
normalized-cut objective has a fractional structure. The authors of
\cite{discriminant_Dias} focus on the two-class clustering and formulate
the optimization objective as a single ratio. In contrast, \cite{Chen_PAMI} 
is concerned with the general multi-class clustering problem which has a
sum-of-ratios optimization objective for which the quadratic transform can be used to decouple the multiple ratios to
facilitate iterative optimization. Other recent applications of FP in
machine learning include the submodular balanced clustering
\cite{submodular_Kawahara} and the fractional loss function approximation
for federated learning \cite{xiaojun_air}.

It is worth pointing out that the practical applications have also conversely
pushed the FP theory forward. Two specific examples follow. First,
owing to the celebrated Shannon's capacity formula
$C=\log(1+\mathrm{SINR})$, the work \cite{shen2018fractional2} proposes a
novel FP technique termed the \emph{Lagrangian dual transform} to address
the log-ratio problem. Second, to account for MIMO transmission, a line of
studies \cite{shen2019optimization,shen2020enhanced,Yannan_TSP24} 
develop a matrix generalization of the traditional scalar-valued FP to
account for the ratio between two matrix-valued functions. Moreover, extensive
connections have been discovered between the FP method and other existing
optimization methods for signal processing and machine learning, e.g., the
fixed-point iteration, the weighted minimum mean squared error (WMMSE)
algorithm, the minorization-maximization or majorization-minimization (MM) method, and the gradient
projection method. In fact,
the latest advances in the FP field mirror the newest frontiers in signal processing. For instance, as opposed to the conventional FP
study that considers the ratio maximization and the ratio minimization
separately, the recent work \cite{Yannan_TSP24} aims at a unified approach
to solve the mixed max-and-min FP problems, as motivated by the emerging
application of ISAC, where the maximization of the SINRs and the
minimization of the CRB coexist. As such, it is worthwhile to look at the
state-of-the-art FP techniques in conjunction with their latest applications. 

The rest of the article is organized according to the classification of the various FP problems. We begin with the single-ratio FP and the max-min-ratios FP, both of which can be solved by the classic methods. Next, we focus on the sum-of-ratios FP. Since the classic methods no longer work for the sum-of-ratios FP, a new method called the quadratic transform is introduced. In particular, the maximization case and the minimization case of the sum-of-ratios FP need to be treated differently. As a generalization of the sum-of-ratios FP, we further discuss the sum-of-functions-of-ratio FP. We then pay special attention to the sum-of-logarithmic-ratios FP, because of the key role it plays in communication system design. Moreover, we discuss the matrix-ratio FP. Lastly, we connect the quadratic transform to other optimization methods and also analyze the rate of convergence.

\emph{Notation:} We denote by $\mathbb R$ the set of real numbers, $\mathbb C$ the set of complex numbers, and $\mathbb S^{m\times m}_{+}$ (resp. $\mathbb S^{m\times m}_{++}$) the set of $m\times m$ positive semi-definite (resp. definite) matrices. We denote by $\|\cdot\|_2$ the Euclidean norm, and $\|\cdot\|_\mathrm{F}$ the Frobenius norm. For a matrix $\bA$, let $\bA^\hh$ be its conjugate transpose and $\bA^\top$ be its transpose. For a square matrix $\bA$, let $\bA^{-1}$ be its inverse (assuming that $\bA$ is nonsingular) and $\mathrm{Tr}(\bA)$ be its trace. Denote the identity matrix by $\bI$. For a real number $a$, let $[a]_+=\max(a,0)$.  Moreover, we use a letter without subscript to denote a set of variables over the subscripts, e.g., $p$ as $\{p_i\}_{i=1}^n$, and $\mathbf Y$ as $\{\mathbf Y_i\}_{i=1}^n$.

\section{Single-Ratio Problem}

Consider a pair of numerator function $A(x)$ and denominator function $B(x)$. Assume that $A(x)\ge0$ and $B(x)>0$ with the variable $x$ restricted to a nonempty set $\mathcal X$. We seek the optimal $x\in\mathcal X$ to maximize an objective that has a ratio form:
\begin{equation}
\label{prob:single ratio max}
\begin{aligned}
\underset{x\in\mathcal X}{\text{maximize}}& \quad \frac{A(x)}{B(x)}.
\end{aligned}
\end{equation}
Unless stated otherwise, we assume by convention that $A(x)$ is concave in $x$, $B(x)$ is convex in $x$, and $\mathcal{X}$ is a nonempty convex set, i.e., the \emph{concave-convex condition} \cite{shen2018fractional}. Note that problem  \eqref{prob:single ratio max} is still nonconvex in general under the concave-convex condition.

The coupling between $A(x)$ and $B(x)$ is the main difficulty in solving problem \eqref{prob:single ratio max}. A natural idea is to decouple the ratio. This constitutes the fundamental basis of most FP methods (including the quadratic transform). For instance, the classic \emph{Charnes-Cooper method} \cite{charnes_1962,schaible_transform} rewrites problem \eqref{prob:single ratio max} as
\begin{equation}
\label{prob:single_ratio_schaible}
\begin{aligned}
\underset{z,\,q}{\text{maximize}} &\quad
z A\left(\frac{q}{z}\right)\\
\text{subject to}
&\quad z B\left(\frac{q}{z}\right)\le1\\
%\label{prob:single_ratio_schaible_constr}\\
&\quad z \in \mathcal{Z},\;q \in \mathcal{Q},
\end{aligned}
\end{equation}
with two new variables introduced as
\begin{equation}
\label{var_zq}
    z = \frac{1}{B(x)} \quad \text{and} \quad q = \frac{x}{B(x)},
\end{equation}
where the constraint sets $\mathcal Z$ and $\mathcal Q$ are determined by going over all $x\in\mathcal X$.
Observe that only $A(x)$ is kept in the new objective function, while $B(x)$ is moved to the constraint. Most importantly, since $A(x)$ is concave and $B(x)$ is convex, the new problem is jointly convex in $(z,q)$, so it can be efficiently solved by standard methods. After obtaining the optimal $(z^\star,q^\star)$, we can immediately recover the optimal solution of the original problem \eqref{prob:single ratio max} as $x^\star=q^\star/z^\star$ according to \eqref{var_zq}.

In comparison, the ratio decoupling technique of the classic \emph{Dinkelbach's method} \cite{dinkelbach_transform} is more straightforward. It breaks down problem \eqref{prob:single ratio max} into a sequence of subproblems
\begin{equation}
\label{prob:Dinkelbach}
\begin{aligned}
\underset{x\in\mathcal X}{\text{maximize}}& \quad A(x) - yB(x),
\end{aligned}
\end{equation}
where the auxiliary variable $y$ is iteratively updated as
\begin{equation}
\label{Dinkelbach:opt y}
    y = \frac{A(x)}{B(x)}.
\end{equation}
Under the concave-convex condition, subproblem \eqref{prob:Dinkelbach} is convex in $x$ and hence can be efficiently solved for fixed $y$. By solving a sequence of subproblems \eqref{prob:Dinkelbach} with $y$ iteratively updated as in \eqref{Dinkelbach:opt y}, Dinkelbach's method guarantees that the sequence of solutions $x$ converges to the optimal solution $x^\star$ in \eqref{prob:single ratio max}. %Furthermore, the rate of convergence is superlinear \cite{dinkelbach_superlinear}.

Thus, both the Charnes-Cooper method and Dinkelbach's method can attain the global optimum of problem \eqref{prob:single ratio max} despite its nonconvexity. This is unsurprising because problem \eqref{prob:single ratio max} is \emph{quasi-convex} (and further \emph{pseudo-convex} \cite{IM_fp}) under the concave-convex condition. The following example shows an early application of the traditional single-ratio FP methods in wireless communications.

\begin{example}[Energy Efficiency of Wireless Link]
\label{example:single-link EE}
Consider a single wireless link. Denote by $p$ the transmit power,  $h\in\mathbb C$ the channel gain, and $\sigma^2$ the noise power. The channel capacity is computed as $\log(1+|h|^2p/\sigma^2)$. Aside from $p$, the wireless transmission system requires a constant {\footnotesize{ON}}-power of $\delta>0$. We seek the optimal $p$ that maximizes the energy efficiency \cite{jorswieck,FettweisTWC2012}:
\begin{equation}
\begin{aligned}
\underset{p}{\text{maximize}}& \quad \frac{\log(1+|h|^2p/\sigma^2)}{p+\delta}\\
\text{subject to}& \quad\, 0\le p\le P,
\end{aligned}
\end{equation}
where $P$ is the max power constraint. Observe that the above single-ratio problem satisfies the concave-convex condition, so both the Charnes-Cooper method and Dinkelbach's method are applicable. 
\end{example}

We remark that not all single-ratio problems in the literature meet the
concave-convex condition.  In the above example, when there are multiple 
wireless links (so each link has its own power variable), the numerator 
becomes the sum rate which is no longer a concave function of the power 
variables, so the new problem after applying Dinkelbach's method is still nonconvex. 
Further optimization steps are required after decoupling the ratio,
as discussed in \cite{jorswieck,shen2018fractional2}.
As another example,
the RTLS problem for image deblurring \cite{Beck_SIAM06_QCFQ} aims to
maximize the ratio between two convex quadratic functions. In this case, as shown in
\cite{jorswieck}, decoupling the ratio (e.g., by Dinkelbach's method) is
still beneficial even when the concave-convex condition does not hold; the
FP technique can be applied in conjunction with other methods to
facilitate optimization.

%When $r$ is increased, the SDP approximation becomes tighter and its solution eventually becomes exactly as the solution of the rational optimization problem at a finite order, (although the complexity of SDP also becomes higher as $r$ increases). 

An important class of the single-ratio FP is the \emph{rational optimization} \cite{Jibetean}, wherein both $A(x)$ and $B(x)$ are polynomial functions of $x\in\mathbb R^d$. The problem is closely related to the polynomial sum of squares and \emph{Hilbert's 17th Problem} \cite{Jibetean}. The state-of-the-art approach in this case is to consider an SDP approximation parameterized by an integer $r\ge0$, based on the epigraph lifting \cite{Jibetean} or the generalized moment problem \cite{Bugarin2015}. When $r$ is increased, the SDP approximation becomes tighter and its solution eventually becomes exactly the solution of the rational optimization problem at a finite order \cite{Nie} (although the complexity of
SDP also becomes higher as $r$ increases). This is known as \emph{Lasserre's hierarchy} \cite{Lasserre2001}. Furthermore, Lasserre's hierarchy can be extended for the sum-of-rational-functions optimization \cite{Bugarin2015}.

\section{Max-Min-Ratios Problem}

We now consider FP problems comprising more than one ratio, writing the numerator and denominator of the $i$th ratio term as $A_i(x)$ and $B_i(x)$, respectively. 
Similar to the single-ratio case, each numerator $A_i(x)$ is assumed to be nonnegative, while each denominator $B_i(x)$ is assumed to be strictly positive. Moreover, the concave-convex condition for the multi-ratio FP now means that each $A_i(x)$ is a concave function, each $B_i(x)$ is a convex function, and $\mathcal X$ is a nonempty convex set. We begin with the max-min-ratios problem:
\begin{equation}
\label{prob:maxmin_ratio}
\begin{aligned}
\underset{x\in\mathcal X}{\text{maximize}} &\quad
\min_{i}\left\{\frac{A_i(x)}{B_i(x)}\right\}.
\end{aligned}
\end{equation}
Like the single-ratio case, the max-min-ratios FP problem is still nonconvex even under the concave-convex condition, so solving it directly is difficult.

As a classic result on the multi-ratio FP \cite{crouzeix1985algorithm}, Dinkelbach's method can be generalized for the max-min-ratios problem by rewriting \eqref{prob:maxmin_ratio} as
\begin{equation}
\label{prob:maxmin_ratio Dinkelbach}
\begin{aligned}
\underset{x\in\mathcal X}{\text{maximize}} &\quad
\min_{i}\left\{A_i(x)-yB_i(x)\right\},
\end{aligned}
\end{equation}
where the auxiliary variable $y$ is iteratively updated as
\begin{equation}
\label{generalized Dinkelbach:y}
    y = \min_{i}\left\{\frac{A_i(x)}{B_i(x)}\right\}.
\end{equation}
It is then possible to reach the global optimum of the original problem \eqref{prob:maxmin_ratio} by solving the new problem \eqref{generalized Dinkelbach:y} with $y$ updated iteratively as above. We now present an application of this technique to a classic problem in machine learning.

\begin{example}[Margin Maximization for SVM]
\label{example:SVM}
\begin{figure}[t]
    \centering
    \vspace{1em}
    \includegraphics[width=0.8\linewidth]{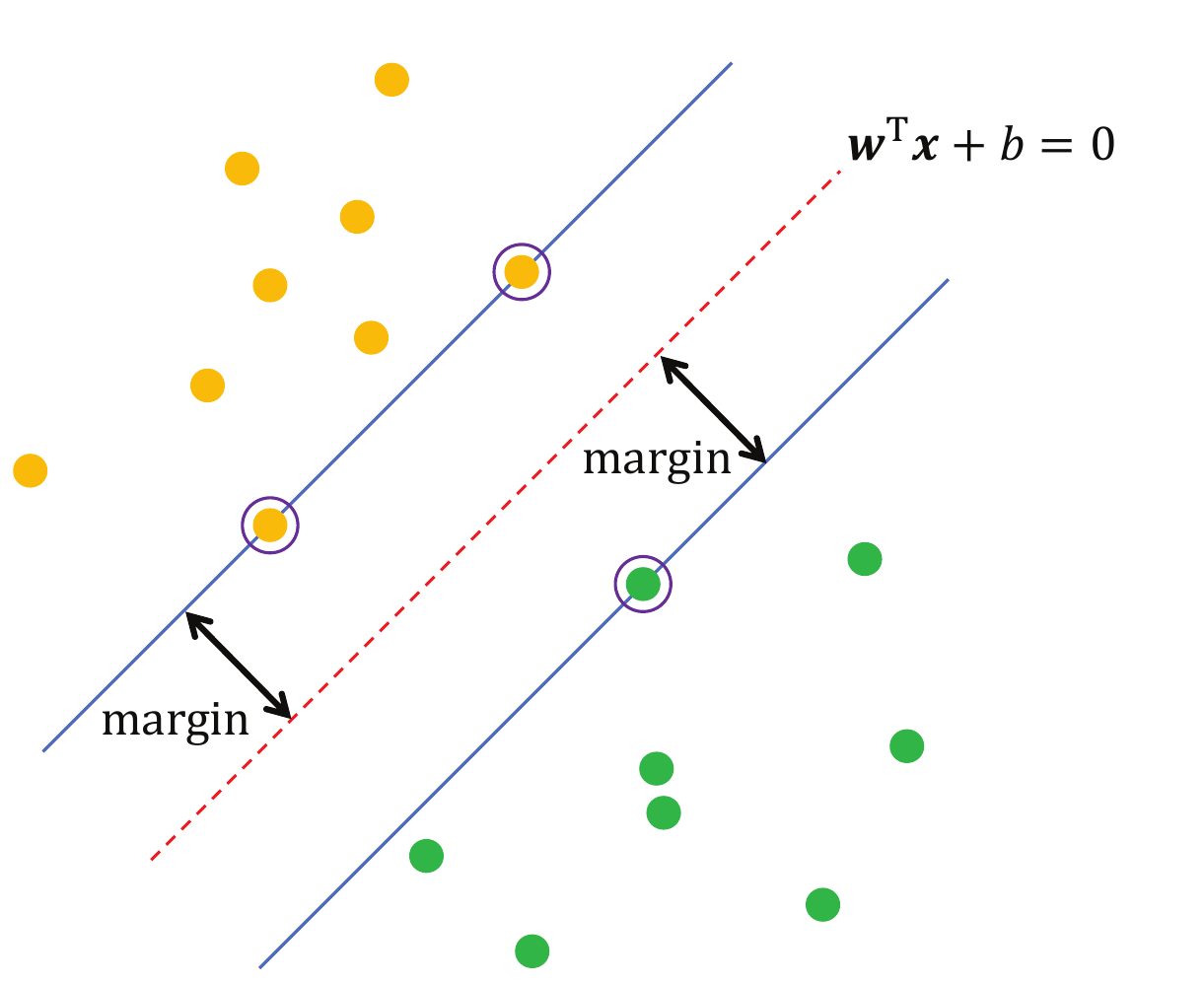}
    \vspace{1em}
    \caption{The yellow points are labeled ``$-1$''; the green points are labeled ``$+1$''. The red dashed line is the decision boundary. The optimization problem of SVM in Example \ref{example:SVM} aims to find the decision boundary to maximize the margin. Because the margin is a fractional function, the optimization in SVM is a max-min-ratios problem.}
    \label{fig:SVM}
\end{figure}
SVM is an important tool for data classification and regression. The core of SVM lies in the minimization of the margin, which turns out to be a max-min-ratios problem. For ease of discussion, consider the binary classification of linearly separable data points. Given a set of data points $\{\bx_1,\bx_2,\ldots,\bx_n\}$, where each $\bx_i\in\mathbb R^m$ has a binary label $t_i\in\{+1,-1\}$, for $i=1,2,\ldots,n$, the aim of the classifier is to draw a decision boundary
$y=\bw^\top\bm\bx+b$
to divide the data points, i.e., label ``$+1$'' if $y>0$ and ``$-1$'' otherwise. The distance from the data point $\bx_i$ to the decision boundary, denoted by $d_i$, can be computed as
\begin{equation}
    d_i = \frac{t_i(\bw^\top\bm\bx_i+b)}{\|\bw\|_2}.
\end{equation}
The shortest distance $d_i$ across all the data points is called the \emph{margin}, as shown in Fig.~\ref{fig:SVM}. Assuming that the data points are separable by a hyperplane, the objective of SVM is to seek the optimal decision boundary that maximizes the margin:
\begin{equation}
%\begin{aligned}
\underset{\bw\in\mathbb R^m,\;b\in\mathbb R}{\text{maximize}} %&
	\quad \min_{i}\{d_i\}. %\\
%\text{subject to}&\quad\, t_i(\bw^\top\bm\bx_i+b)>0,\quad i=1,\ldots,n,
%\end{aligned}
\end{equation}
The objective function here is a ratio, which is nonconvex. The usual 
treatment in most standard textbooks on this subject proceeds to transform
the problem into an equivalent convex quadratic form. 
Subsequently, Lagrangian duality theory is applied to solve the problem in
the dual domain. 

Here, we point out that since each $d_i$ is a fractional function, problem \eqref{prob:maxmin_ratio} can be directly treated as a max-min-ratios problem. In particular, \eqref{prob:maxmin_ratio} satisfies the concave-convex condition, so it can be immediately solved by the generalized Dinkelbach's method \cite{crouzeix1985algorithm}. Aside from using this max-min-ratios FP approach, we can alternatively regard the entire $\min_i\{t_i(\bw^\top\bm\bx_i+b)\}$ as the numerator and notice that the resulting single-ratio problem meets the concave-convex condition, so the Charnes-Cooper method \cite{charnes_1962,schaible_transform} and Dinkelbach's method \cite{dinkelbach_transform} are also applicable here.% Furthermore, for the multi-class SVMs involving multiple margins, the multi-ratio FP comes into play. 

We remark here that the max-min-ratios FP is also involved in the so-called Grab-n-Pull framework with broad applications for relay beamforming, Doppler-robust waveform design for active sensing, and robust data classification \cite{OtterstenSP2019robustClustering}.
\end{example}

\section{Sum-of-Ratios Problem}
\label{subsec:sum of weighted ratios}

As compared to the max-min-ratios problem, the sum-of-ratios problems are
much more challenging. It is known
\cite{freund} that both the maximization and minimization of the sum of
ratios are NP-complete, even when the numerators and denominators are all
linear functions. For example, a sum-of-ratios problem with $20$ ratios
already cannot be globally solved within reasonable time according to \cite{kuno}.
As such, finding a local optimum efficiently is what one can expect at
best.

The difference between maximization and minimization adds a further
complication to the sum-of-ratios problems. The previous section only 
discusses how to maximize a single ratio as in \eqref{prob:single ratio max} or how to maximize min-ratios as in 
\eqref{prob:maxmin_ratio}. If instead we consider the minimization of a
single-ratio or the minimization of max-ratios, it would suffice to take 
the reciprocal(s) of the ratio term(s) and apply the
same technique as before (with the concave-convex condition reversed as 
the convex-concave condition).  However, when it comes to the sum-of-ratios
problem, we can no longer convert the minimization to maximization by
simply taking the reciprocals. For this reason, we discuss the
sum-of-ratios maximization problem and the sum-of-ratios minimization 
problem separately.

\subsection{Sum-of-Ratios Maximization Problem}

We start with the sum-of-ratios maximization problem:
\begin{equation}
\label{prob:sum of ratios}
\begin{aligned}
\underset{x\in\mathcal X}{\text{maximize}}& \quad \sum^n_{i=1}\frac{A_i(x)}{B_i(x)}.
\end{aligned}
\end{equation}
The natural extension of Dinkelbach's method from the single-ratio to the
max-min-ratios problem as discussed earlier may lead one into believing
that other multi-ratio problems can be dealt with in a similar fashion.
Unfortunately, the max-min-ratios problem is a rare case, whereas most
multi-ratio problems cannot be addressed by extending Dinkelbach's 
method. For instance, it is tempting to generalize Dinkelbach's method
for the above sum-of-ratios problem by using auxiliary variables $y_i$
updated as $A_i(x)/B_i(x)$ in an iterative fashion, then maximizing a 
transformed problem in each step as in the single-ratio Dinkelbach's 
method. But this 
is not equivalent to the original problem \eqref{prob:sum of ratios}. 
In other words,
\begin{equation}
%\exists ! \ y_i \ {\mathrm{s.t.}}\ 
\underset{x \in \mathcal X}{\text{maximize}}\; \sum^n_{i=1}\frac{A_i(x)}{B_i(x)} \notleftright
\underset{x \in \mathcal X}{\text{maximize}}\; \sum^n_{i=1}\Big({A_i(x)}-y_i{B_i(x)}\Big),
\end{equation}
when each $y_i$ is iteratively updated as $y_i=A_i(x)/B_i(x)$. As explained in \cite{shen2018fractional}, the fundamental reason for the
breakdown of Dinkelbach's method in the multi-ratio case is that the 
problem transformation lacks the \emph{objective value equivalence}. 
This is to say that, although problem \eqref{prob:single ratio max} and 
problem \eqref{prob:Dinkelbach} have the same solution set, their 
objective values are not equal at the optimum. Thus, one cannot add 
multiple single-ratios together and expect the transformed problem to be
equivalent to the original problem. For specific problems, e.g., energy efficiency problem, \cite{OtterstenTSP2014EE} suggests rewriting the sum-of-ratios problem as a parameterized polynomial optimization problem, while \cite{ZapponeTWC2015} proposes a water-filling-type algorithm.

To fix this issue in general, we propose a new ratio-decoupling transform that imposes the objective
value equivalence, i.e., the objective value of the new problem must be
equal to that of the original problem at the optimum. Moreover, we require
the new objective function to be concave in the auxiliary variable for ease
of iterative update. Under the above two assumptions, it is shown in
\cite{shen2018fractional} that the new transform can take a quadratic
function form. Specifically, problem \eqref{prob:sum of ratios} can be
rewritten as
\begin{equation}
\label{prob:quadratic transform}
\begin{aligned}
\underset{x,\,  y}{\text{maximize}}& \quad \sum^n_{i=1}\Big(2y_i\sqrt{A_i(x)}-y^2_iB_i(x)\Big).
\end{aligned}
\end{equation}
This new transform is termed \emph{quadratic transform} as first
proposed in \cite{shen2018fractional}. We  discuss why the quadratic form is preferable to other function forms later in this article when discussing the connection between the quadratic transform and the MM method. The new problem \eqref{prob:quadratic transform} is amenable to alternating optimization. When $x$ is held fixed, each $y_i$ is optimally determined as
\begin{equation}
\label{QT:opt y}
    y^\star_i = \frac{\sqrt{A_i(x)}}{B_i(x)}.
\end{equation}
When $  y$ is held fixed, solving for $x$ in \eqref{prob:quadratic transform} is a convex problem under the concave-convex condition. Later in the article, we see that there are also applications in which the concave-convex condition does not hold, yet $x$ can still be efficiently optimized for fixed $y$. Under the concave-convex condition (and also assuming differentiability of $A_i(x)$ and $B_i(x)$), the iterative optimization over $x$ and $y$ is guaranteed to converge to a stationary point of the original optimization problem \eqref{prob:sum of ratios}.

An intuitive comparison between Dinkelbach's method
\cite{dinkelbach_transform} and the quadratic transform
\cite{shen2018fractional} is as follows. 
To maximize a ratio $A/B$, we need to increase
$A$ and decrease $B$ simultaneously. Thus, $A$ can be treated as the
utility and $B$ the penalty. Dinkelbach's method, $A-yB$, then boils down
to a utility-minus-penalty strategy, where the auxiliary variable $y$
serves as the price for the penalty. The quadratic transform method,
$2y\sqrt{A}-y^2B$, combines the utility and penalty differently. Because
the quadratic transform places $A$ inside a concave operation (i.e., the square root), it is
less aggressive than Dinkelbach's method in boosting $A$. As a consequence,
the quadratic transform converges more slowly than Dinkelbach's method when
solving a single-ratio problem, but it enjoys an advantage in that
it can be applied to multi-ratio FP, whereas Dinkelbach's method cannot.

\subsection{Sum-of-Ratios Minimization Problem}
\label{subsec: sum of ratios min}

We proceed to the case of minimization of the sum-of-ratios problem, written in a slightly different form as below:
\begin{equation}
\label{prob:sum of ratios min}
\begin{aligned}
\underset{x\in\mathcal X}{\text{maximize}}& \quad -\sum^n_{i=1}\frac{A_i(x)}{B_i(x)}.
\end{aligned}
\end{equation}
We write the minimization problem in the maximization form with the
optimization objective multiplied by $-1$, in order to simplify the
notation in the later developments.  As already mentioned before, instead 
of the concave-convex condition in the maximization case, a common 
condition adopted for the minimization problem is that each
$A_i(x)$ is a convex function, each $B_i(x)$ is a concave function, and
$\mathcal X$ is still a nonempty convex set, namely the
\emph{convex-concave condition}. A naive idea is to merge $-1$ into
$A_i(x)$ or $B_i(x)$ and thereby apply the previous FP method for the
maximization problem; this is however
problematic since the resulting new FP problem violates the condition
that each numerator is nonnegative while each denominator is positive. 
Two methods emerged in the recent literature to handle the sum-of-ratios 
minimization problem.

The first method \cite{Yannan_TSP24} aims to extend the quadratic
transform to what is called the \emph{inverse quadratic transform}. 
It recasts the original problem \eqref{prob:sum of ratios min} as
\begin{equation}
\label{prob:FP_min:quadratic}
\begin{aligned}
\underset{x\in\mathcal X,\,  y}{\text{maximize}}& \quad - \sum^n_{i=1}\frac{1}{\Big[2 y_i\sqrt{B_i(x)}-y_i^2A_i(x)\Big]_+}.
\end{aligned}
\end{equation}
Note that the denominator is placed inside $[\,\cdot\,]_+$ in \eqref{prob:FP_min:quadratic} in order to rule out negative value in the denominator. This operation is critical to the equivalence between the problems \eqref{prob:sum of ratios min} and \eqref{prob:FP_min:quadratic}, for otherwise letting $y_i\uparrow0$ while fixing $x$ would make the objective in \eqref{prob:FP_min:quadratic} go to infinity, thus making the maximization problem degenerate. For the new problem \eqref{prob:FP_min:quadratic}, we again optimize $x$ and $  y$ in an alternating fashion, which is guaranteed to reach a stationary point of the original problem. When $x$ is held fixed, each $y_i$ can be optimally updated as
\begin{equation}
    y^\star_i = \frac{\sqrt{B_i(x)}}{A_i(x)}.
\end{equation}
For fixed $  y$, optimizing $x$ in \eqref{prob:FP_min:quadratic} is a convex problem under the convex-concave condition.

Another method, as recently proposed in \cite{ZhaoJun_JSAC}, is called the arithmetic-mean geometric-mean (AM-GM) inequality transform. It rewrites problem \eqref{prob:sum of ratios min} as
\begin{equation}
\label{prob:AM-GM}
\begin{aligned}
\underset{x\in\mathcal X,\,  y}{\text{maximize}}& \quad -\sum^n_{i=1}\bigg(y_iA^2_i(x)+\frac{1}{4y_iB^2_i(x)}\bigg).
\end{aligned}
\end{equation}
Like \eqref{prob:FP_min:quadratic}, the above new problem is amenable to the alternating optimization between $x$ and $  y$ to reach a stationary point. Specifically, for fixed $x$, each $y_i$ is optimally updated as
\begin{equation}
\label{Zhao-Qian-Yu opt y}
    y^\star_i = \frac{1}{2A_i(x)B_i(x)}.
\end{equation}
For fixed $  y$, problem \eqref{prob:AM-GM} is convex in $x$ under the convex-concave condition, so it can be solved efficiently. 

Although the inverse quadratic transform \cite{Yannan_TSP24} and the AM-GM inequality transform \cite{ZhaoJun_JSAC} look quite different, it turns out that both of them can be placed under the umbrella of the MM theory \cite{razaviyayn2013unified,sun2016majorization}, so their convergence can be verified immediately. A later section is dedicated to this connection. We now present an application of the sum-of-ratios minimization FP.

\begin{example}[Age-of-Information (AoI) Minimization in Queuing System]
\label{example:AoI}
\begin{figure}[t]
\centering
\subfigure[A $K$-source queuing system with priority.]{
\includegraphics[width=0.6\linewidth]{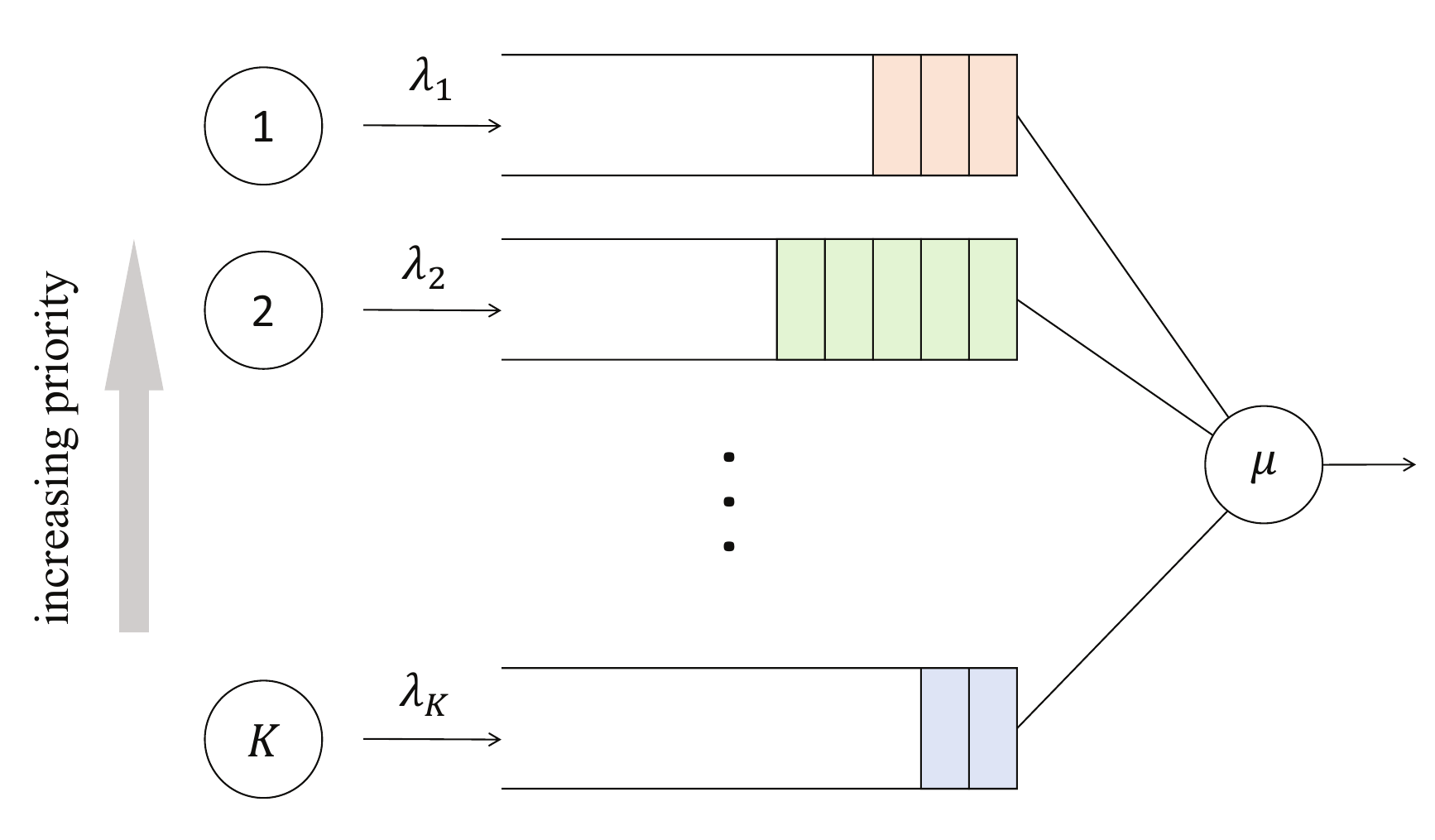}}\\
\hspace{1em}
\subfigure[A typical instantaneous AoI $\Delta_k$ curve versus time.]{
\includegraphics[width=0.7\linewidth]{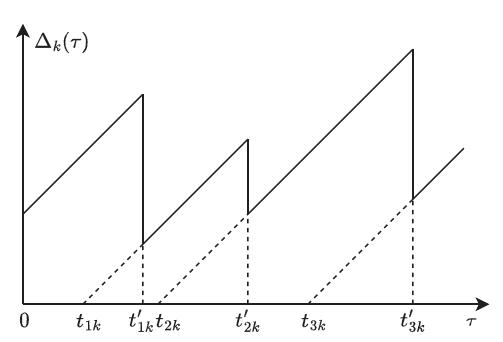}}
\caption{Rate control for minimizing AoI in Example \ref{example:AoI}. The average AoI $\bar\Delta_k$ is the average area of the trapezoid below each tooth of the sawtooth curve. Since $\bar\Delta_k$ is a sum of ratios, the problem of optimizing the sensor data packet arrival rate to minimize the average AoI is a sum-of-ratios minimization problem.}
\label{fig:diagram}
\end{figure}
The notion of AoI \cite{Yates_JSAC21} characterizes the freshness of data packet. Consider a $K$-sensor queuing network as depicted in Fig.~\ref{fig:diagram}(a). Each sensor $k$ samples its source and delivers data packets at a rate $\lambda_k$ (which is to be optimized), while the server processes the received packets at a constant rate $\mu$, but treats the sensors with lower indices at higher priority, i.e., the packets from sensor $k$ are served only if the queues associated with sensors $1, \cdots, k-1$ are all empty. The AoI for each sensor is defined as follows. The $i$th packet from the sensor $k$ is delivered at time $t_{ik}$ and departs the server at time $t'_{ik}$; the delay $t'_{ik}-t_{ik}$ is due to the waiting time in the queue and the server processing time. At the current time $\tau$, let $\mathcal N_k(\tau)$ be the arrival time of the most recently received packet from the sensor $k$, i.e., $\mathcal N_k(\tau) = \max\left\{t'_{ik}:t'_{ik}\le \tau\right\}$. The instantaneous AoI of the source $k$ at time $\tau$ is given by $\Delta_k(\tau) = \tau - \mathcal N_k(\tau)$. As a result, $\Delta_k(\tau)$ increases linearly with $\tau$, and drops whenever a new packet departs the server, so $\Delta_{k}(\tau)$ has a sawtooth profile along the time axis as shown in Fig.~\ref{fig:diagram}(b). We are interested in the average AoI in the long run $\mathbb E[\Delta_k] = \lim_{T\rightarrow\infty}\frac{1}{T}\int_{0}^{T}\Delta_k(\tau)d\tau$, which is the average area of the trapezoid below each tooth of the sawtooth curve as in Fig.~\ref{fig:diagram}(b). For the M/M/1 queue model \cite{Yates_JSAC21}, the problem of minimizing the
sum-of-AoI for the $K$ sources can be formulated
as
\begin{equation}
\begin{aligned}
\underset{ \lambda}{\text{minimize}} &\quad \sum_{k=1}^K \bigg(\frac{\hat{\rho}_{k}^{2}+3\hat{\rho}_{k}+1}{\mu(1+\hat{\rho}_k)}+\frac{(\hat{\rho}_k+1)^2}{\mu\rho_k}\bigg) \\
\text{subject to}
&\quad\; 0\le\lambda_k\le\mu,\quad k=1,\ldots,K,
\end{aligned}
\end{equation}
where $\rho_k= {\lambda_k}/{\mu}$ and $\hat{\rho}_{k} = \sum_{i=1}^{k-1}\rho_i$. This problem can be recognized as a sum-of-ratios minimization problem satisfying the convex-concave condition, so we can apply either the inverse quadratic transform \cite{Yannan_TSP24} or the AM-GM inequality transform \cite{ZhaoJun_JSAC}. A numerical example is shown in Fig.~\ref{fig:AoI different K}, where the processing rate is fixed at $\mu=1$ and the numbers of source nodes varies as $K=3,4,\ldots,10$. Two benchmarks are included in the figure: 1) \emph{equal-rate optimization} \cite{kaul2018age}, which assumes all $\lambda_k$'s are equal and then performs a one-dimensional search, and 2) \emph{max-rate scheme}, which sets each $\lambda_k=\mu$.
Observe that the FP method achieves a much lower sum of AoIs and thus provides fresher information. Moreover, we remark that the AoI problem can be also tackled by the extended Lasserre's hierarchy \cite{Bugarin2015}, since the numerators and denominators are all polynomials.
\begin{figure}[t]
  \centering
  \includegraphics[width=0.95\linewidth]{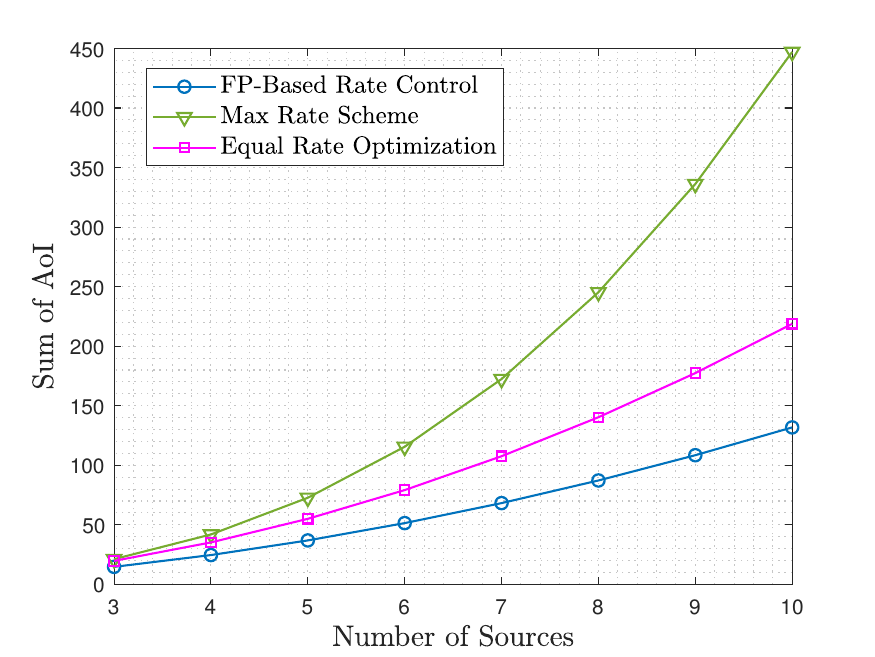}
  \caption{Minimization of the sum of AoIs by different methods in Example \ref{example:AoI}.}
  \label{fig:AoI different K}
\end{figure}
\end{example}

\section{Sum-of-Functions-of-Ratio Problem}
\label{subsec:mixed max and min}

We now consider a further extension of the sum-of-ratios optimization by investigating mixed functions of ratios. To this end, consider a sequence of nondecreasing functions $f^+_i:\mathbb R_+\rightarrow\mathbb R$ and a sequence of nonincreasing functions $f^-_i:\mathbb R_+\rightarrow\mathbb R$, each of which takes a ratio term ${A(x)}/{B(x)}$ as input. We then arrive at a generalized sum-of-functions-of-ratio problem:
\begin{equation}
\label{prob:FP_mixed}
\begin{aligned}
%\underset{x}{\text{maximize}}& \quad \sum^n_{i=1}\bigg[f^+_i\left(\frac{A_i(x)}{B_i(x)}\right)+f^-_i\left(\frac{A_i(x)}{B_i(x)}\right)\bigg]\\
\underset{x\in\mathcal X}{\text{maximize}}& \;\; 
\sum^{n^+}_{i=1} f^+_i\left(\frac{A^+_i(x)}{B^+_i(x)}\right) + 
\sum^{n^-}_{i=1} f^-_i\left(\frac{A^-_i(x)}{B^-_i(x)}\right).
\end{aligned}
\end{equation}
Intuitively, we aim to maximize the ratios inside $f^+_i(\,\cdot\,)$ and minimize the ratios inside $f^-_i(\,\cdot\,)$ at the same time. 
Thus, we can apply the quadratic transform to every ratio inside $f^+_i(\,\cdot\,)$
and in the meantime apply the inverse quadratic transform to every ratio
inside $f^-_i(\,\cdot\,)$, thereby converting problem \eqref{prob:FP_mixed}
to
\begin{equation}
\label{prob:FP_mixed:quadratic}
\begin{aligned}
\underset{x\in\mathcal X,\,  y,\, {\tilde y}}{\text{maximize}} &\quad \sum^{n^+}_{i=1}f^+_i\left(2y_i\sqrt{A^+_i(x)}-y^2_iB^+_i(x)\right)\\
&\quad+\sum^{n^-}_{i=1}
f^-_i\left({\Big[2\tilde y_i\sqrt{B^-_i(x)}-\tilde y_i^2A^-_i(x)\Big]_+^{-1}}\right),
\end{aligned}    
\end{equation}
where the auxiliary variables introduced by the quadratic transform are denoted by $y_i$, and the auxiliary variables introduced by the inverse quadratic transform are denoted by $\tilde y_i$. The preceding generalized quadratic transform is referred to as the \emph{unified quadratic transform}. 

As shown in a later part of the article, the unified quadratic transform is still an MM procedure, so the alternating optimization continues to guarantee convergence. Furthermore, the alternating optimization can be carried out efficiently provided that each ratio inside $f^+_i(\,\cdot\,)$ meets the concave-convex condition, while each ratio inside $f^-_i(\,\cdot\,)$ meets the convex-concave condition. When $x$ is held fixed, all the auxiliary variables can be simultaneously optimally determined as
\begin{equation}
    y^\star_i=\frac{\sqrt{A^+_i(x)}}{B^+_i(x)} \quad\text{and}\quad \tilde y^\star_i=\frac{\sqrt{B^-_i(x)}}{A^-_i(x)}.
\end{equation}
When the auxiliary variables are held fixed, solving for $x$ in \eqref{prob:FP_mixed:quadratic} is a convex optimization problem. The following is an example of mixed max-and-min problem. It also shows that the choices of $f^+_i(\,\cdot\,)$ and $f^-_i(\,\cdot\,)$ are not necessarily unique for the same problem, and such choices may even be critical to the performance of the unified quadratic transform.

\begin{example}[Secure Transmission]
\label{example:secure transmission}
Consider the downlink of a wireless network with $L$ cells. Within each cell, the base station (BS) transmits data to one legitimate user, at the risk of being wiretapped by an eavesdropper. Use $i=1,2,\ldots,L$ to index each cell and its associated legitimate user and eavesdropper. Denote by $p_i$ the transmit power of BS $i$, $h_{ji}\in\mathbb C$ the channel from BS $i$ to legitimate receiver $j$, $\tilde h_{ji}\in\mathbb C$ the channel from BS $i$ to eavesdropper $j$, $\sigma^2_i$ the noise power at legitimate receiver $i$, and $\tilde\sigma^2_i$ the noise power at eavesdropper $i$. The SINRs of legitimate receiver $i$ and eavesdropper $i$ are respectively given by
\begin{equation}
\Gamma_i=\frac{|h_{ii}|^2p_i}{\sum_{j\ne i}|h_{ij}|^2p_j+\sigma^2_i}\quad\text{and}\quad\widetilde\Gamma_i=\frac{|\tilde h_{ii}|^2p_i}{\sum_{j\ne i}|\tilde h_{ij}|^2p_j+\tilde\sigma^2_i}.
\end{equation}
We seek the optimal power allocation $  p$ that maximizes the total secrecy data rates:
\begin{equation}
  \label{prob:secure transmission}
\begin{aligned}
\underset{  p}{\text{maximize}} &\quad \sum^{L}_{i=1}\Big(\log\big(1+\Gamma_i\big)-\log\big(1+\widetilde\Gamma_i\big)\Big)\\
\text{subject to}&\quad\, 0\le p_i\le P,\quad i=1,\ldots,L, 
\end{aligned}    
\end{equation}
where $P$ is the power constraint on each BS. At first glance, the unified quadratic transform seems to be immediately applicable for problem \eqref{prob:secure transmission} by letting $f^+_i(r)=\log(1+r)$ and $f^-_i(r)=-\log(1+r)$. But such $f^-_i(r)$ is not a concave function, so it is difficult to optimize $  p$ in the new problem when the auxiliary variables are fixed. The above issue can be resolved by rewriting the secrecy data rate as
\begin{equation}
  \label{prob:secure transmission new}
\begin{aligned}
\underset{  p}{\text{maximize}} &\quad\; \sum^{L}_{i=1}\left(\log\big(1+\Gamma_i\big)+\log\left(1-\big(1+\widetilde\Gamma_i^{-1}\big)^{-1}\right)\right)\\
\text{subject to}&\quad\;\, 0\le p_i\le P,\quad  i=1,\ldots,L,
\end{aligned}    
\end{equation}
with $f^+_i(r)=\log(1+r)$ and $f^-_i(r)=\log(1-r)$. Observe that the new problem is convex in $  p$ for fixed $\{y_i,\tilde y_i\}$. 

\begin{figure}
\centering
\centerline{\includegraphics[width=0.95\linewidth]{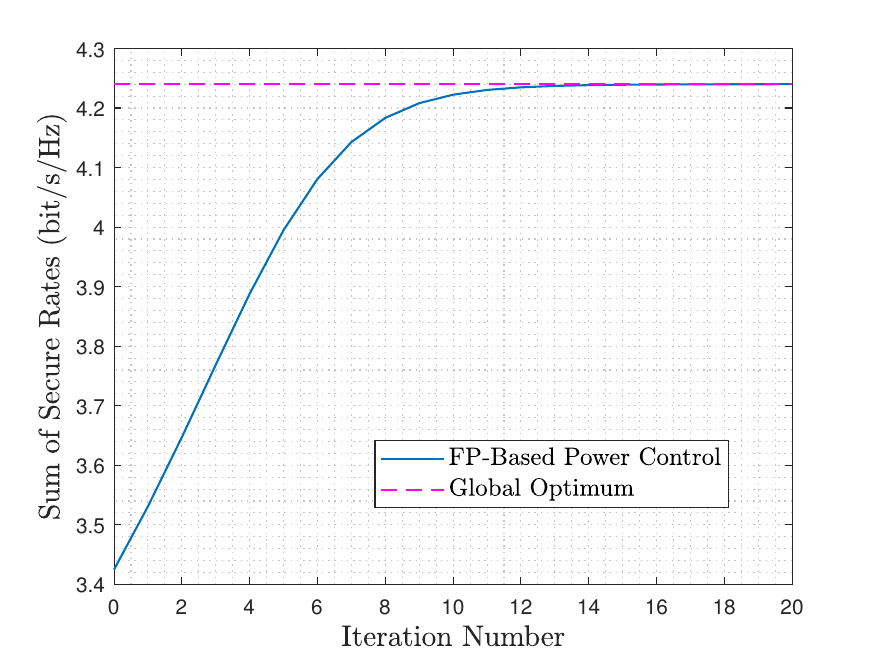}}
\caption{Maximization of the sum of secrecy rates in Example \ref{example:secure transmission}.}
\label{fig:max_secure_rate_iteration}
\end{figure}

Fig.~\ref{fig:max_secure_rate_iteration} shows some simulation results for this example. In this simulation, we consider $L=2$ cells. Let $\sigma_i^2=-10$ dBm, $\tilde\sigma_i^2=0$ dBm, $P=10$ dBm, $|h_{11}|^2=1$, $|h_{12}|^2=0.1$, $|h_{21}|^2=0.09$, $|h_{22}|^2=0.87$, $|\tilde{h}_{11}|^2=0.5$, $|\tilde{h}_{12}|^2=0.11$, $|\tilde{h}_{21}|^2=0.13$, and $|\tilde{h}_{22}|^2=0.39$. The global optimum is obtained via exhaustive search. Observe from Fig.~\ref{fig:max_secure_rate_iteration} that the FP-based power control algorithm converges to the global optimum solution in this example. Observe also that the FP method has fast convergence.
\end{example}

\section{Sum-of-Logarithmic-Ratios Problem}

Because of Shannon's capacity formula for the Gaussian channel, the following log-ratio problem deserves special attention:
\begin{equation}
\label{prob:log ratio}
\begin{aligned}
\underset{x\in\mathcal X}{\text{maximize}}& \quad \sum^n_{i=1}w_i\log\bigg(1+\frac{A_i(x)}{B_i(x)}\bigg),
\end{aligned}
\end{equation}
where each ratio $A_i(x)/B_i(x)$ can be interpreted as the SINR of link $i$, and each positive weight $w_i>0$ can be interpreted as the priority for each link.

The above problem aims to maximize a sum of weighted rates across multiple links. One can immediately recognize problem \eqref{prob:log ratio} as a special case of problem \eqref{prob:FP_mixed}, so the unified quadratic transform is applicable here. As a result, we can solve a sequence of convex subproblems in $x$ with the auxiliary variables iteratively updated. 

However, the above method has two downsides. First, although the unified quadratic transform can convert the log-ratio problem \eqref{prob:log ratio} into a sequence of convex subproblems in $x$, solving for $x$ in each subproblem can only be done as a generic convex optimization problem as shown in the previous example in secure transmission. It would have been more desirable if it is possible to exploit the structure of these subproblems.
Second, if we seek further extension of the FP technique to the case where the constraint set $\mathcal X$ is discrete, then decoupling ratios inside the logarithms does not help much, because we still face a challenging nonlinear discrete optimization problem after the transform. 

Since the nonlinearity of logarithm is the reason for the above issues, a natural idea is to try to ``move'' ratios to the outside of the logarithms. Toward this end, the following \emph{Lagrangian dual transform} has been developed in \cite{shen2018fractional2}, where it is shown that \eqref{prob:log ratio} is equivalent to
\begin{equation}
\label{prob:log ratio:Lagrangian}
\begin{aligned}
\underset{x\in\mathcal X,\, \gamma}{\text{maximize}}& \quad \sum^n_{i=1}w_i\left(\log\big(1+\gamma_i\big)+\frac{(1+\gamma_i)A_i(x)}{A_i(x)+B_i(x)}-\gamma_i\right). 
\end{aligned}
\end{equation}
Here, an auxiliary variable $\gamma_i$ is introduced for each ratio term $A_i(x)/B_i(x)$.
Note that when moving $A_i(x)/B_i(x)$ to outside of the logarithm, we also need to add $A_i(x)$ to the denominator. For fixed $x$, each $\gamma_i$ is optimally determined as 
\begin{equation}
\label{opt gamma}
    \gamma^\star_i = \frac{A_i(x)}{B_i(x)}.
\end{equation}
For fixed $ \gamma$, optimizing $x$ in problem \eqref{prob:log ratio:Lagrangian} boils down to a sum-of-weighted-ratios problem as formerly discussed. The next example shows that using the Lagrangian dual transform coupled with the quadratic transform can result in a sequence of subproblems with the sum-of-ratios structure, whereas using the quadratic transform alone cannot.

\begin{example}[Power Control for Interfering Links]
\label{example:power control}
Consider $K$ wireless links that reuse the same spectral band. Denote by $h_{ij}\in\mathbb C$ the channel from the transmitter of link $j$ to the receiver of link $i$, $p_i$ the transmit power of link $i$, $P$ the power constraint, and $\sigma^2$ the background noise power. We consider the following power control problem of maximizing the sum of weighted rates:
\begin{equation}
%\label{prob:power control}
\begin{aligned}
\underset{  p}{\text{maximize}}& \quad \sum^K_{i=1}w_i\log\Bigg(1+\frac{|h_{ii}|^2p_i}{\sum_{j\ne i}|h_{ij}|^2p_j+\sigma^2}\Bigg)\\
\text{subject to}& \quad\, 0\le p_i \le P,\quad i=1,\ldots,K,
\end{aligned}
\end{equation}   
where the weight $w_i>0$ reflects the priority of link $i$. Although the unified quadratic transform enables a convex reformulation of the power control problem, updating $  p$ still requires solving a generic convex optimization problem.

We now show that using the Lagrangian dual transform coupled with the quadratic transform can lead to an FP problem in each iterate. First, by the Lagrangian dual transform, we move the ratios out of the logarithms:
\begin{equation}
\begin{aligned}
\underset{  p,\, \gamma}{\text{maximize}}& \; \sum^K_{i=1}w_i\Bigg(\log(1+\gamma_i)+\frac{(1+\gamma_i)|h_{ii}|^2p_i}{\sum^K_{j=1}|h_{ij}|^2p_j+\sigma^2}-\gamma_i\Bigg)\\
\text{subject to}& \;\, 0\le p_i \le P,\quad i=1,\ldots,K.
\end{aligned}
\end{equation}
After $ \gamma_i$'s are optimally updated as in \eqref{opt gamma}, solving for $  p$ in the above problem boils down to a sum-of-ratios maximization FP as considered previously in this article. As a result, each $p_i$ can be optimally solved in closed form in each iteration as
\begin{equation}
p_i = \min\left\{P,\,\frac{y^2_iw_i(1+\gamma_i)|h_{ii}|^2}{\big(\sum^K_{j=1}y^2_j|h_{ji}|^2\big)^2}\right\},
\end{equation}
where each $y_i$ is an auxiliary variable introduced by the quadratic transform and is iteratively updated as \begin{equation}y_i=\frac{\sqrt{w_i(1+\gamma_i)|h_{ii}|^2p_i}}{\sum^K_{j=1}|h_{ij}|^2p_j+\sigma^2}.
\end{equation}

\begin{figure}[t]
\centering
\centerline{\includegraphics[width=0.95\linewidth]{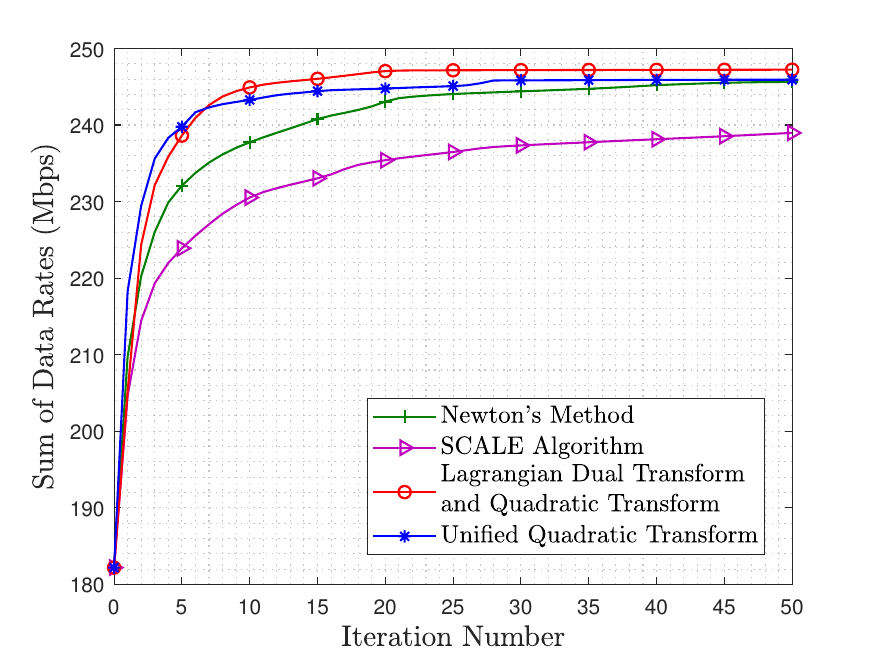}}
\caption{Cumulative distributions of the sum rates in a 7-cell network achieved by the different power control methods discussed in Example \ref{example:power control}.}
\label{fig:power_rate}
\end{figure}

Fig.~\ref{fig:power_rate} validates the performance advantage of the proposed FP-based power control method in a 7-cell wrapped-around network. We aim to maximize the sum of downlink data rates. The BS-to-BS distance is 0.8 km, $P=43$ dBm, $\sigma^2=-100$ dBm, and the spectrum bandwidth is 10 MHz. The pathloss is modeled as $128.1 + 37.6\log_{10} (d) + \tau$ (in dB), where the distance $d$ is in km, and the shadowing $\tau$ has a zero-mean Gaussian distribution with 8 dB standard deviation. To avoid the bias caused by the starting point, we try out the same set of random starting points for the different algorithms and average their performance. We use Newton's method and the geometric programming (GP) based method called the \emph{successive convex approximation for low-complexity (SCALE)} \cite{john} as benchmarks. Observe that the two FP methods achieve competitive sum rates as the benchmarks, but the computational complexity of FP is much lower, because the updates in FP are in closed form.
\end{example}

\section{Matrix-Ratio Problems}
\label{subsec:matrix ratio}

A key advantage of the FP formulation is that it can be generalized to deal with matrix ratios. While the
traditional setup of FP, as discussed thus far in the article, deals with a scalar 
ratio between a nonnegative function $A_i(x)\ge0$ and a strictly positive
function $B_i(x)>0$, we now generalize it to a ratio between a positive
semi-definite function $\bA_i(x)\in\mathbb S^{m\times m}_+$ and a positive
definite function $\bB_i(x)\in\mathbb S^{m\times m}_{++}$. To lighten the
notation, we drop the argument $x$ in these matrix-valued functions in the
rest of the article whenever doing so causes no confusion. Furthermore, we assume that
each matrix numerator $\bA_i$ can be factorized as
$\bA_i = \sqrt{\bA_i}\sqrt{\bA_i}^{\hh}$ where $\sqrt{\bA_i}\in\mathbb C^{m\times \ell}$,  
for some given positive integer $\ell$. Note that the symbol $\sqrt{\bA_i}$ denotes a matrix square root rather than an operation; note also that the above factorizations always exist so long as $\ell\ge \max_i\{\mathrm{rank}(\bA_i)\}$. The traditional scalar-valued ratio is then generalized to the multidimensional case as below: 
\begin{equation}
%\label{matrix ratio}
\frac{A_i}{B_i}\in\mathbb R\quad
\Rightarrow\quad\sqrt{\bA_i}^\hh\bB^{-1}_i\sqrt{\bA_i}\in\mathbb R^{\ell\times \ell}.
\end{equation}
One of the most intriguing aspects of the quadratic transform is that it can be extended to matrix ratios. Below we show how the extension works for the quadratic transform in \eqref{prob:quadratic transform}.

To start, we write the matrix-ratio analog of the
sum-of-ratios problem \eqref{prob:sum of ratios}:
\begin{align}
 \underset{x\in\mathcal X}{\text{maximize}} &\quad
\sum_{i=1}^{n}
\mathrm{Tr}\Big(\sqrt{\bA_i}^\hh\bB_i^{-1}\sqrt{\bA_i}\Big).\label{prob:FP_mixed_matrix}
\end{align}
The matrix ratios can be decoupled by a matrix extension of the quadratic transform 
\begin{equation}
\label{prob:FP_mixed_matrix:quadratic}
\begin{aligned}
\underset{x\in\mathcal X,\, \bY}{\text{maximize}} &\quad\sum_{i=1}^{n}\mathrm{Tr}\Big(\sqrt{\bA_i}^{\hh} \bY_i + \bY^{\hh}_i\sqrt{\bA_i}- \bY_i^{\hh} \bB_i \bY_i\Big)\\
\text{subject to}&\quad\,\bY_i\in\mathbb C^{m\times \ell},\quad i=1,\ldots,n.
\end{aligned}
\end{equation}
The above matrix extension of the quadratic transform preserves the connection to the MM method, i.e., the alternating optimization between $x$ and $\bY$ can still be recognized as an MM procedure, so the iterative process must converge. When $x$ is held fixed, the auxiliary variables are optimally determined as
\begin{equation}
\label{opt Y matrix FP}
    \bY^\star_i=\bB_i^{-1}\sqrt{\bA_i}.
\end{equation}
For fixed $\bY$, solving for $x$ in
\eqref{prob:FP_mixed_matrix:quadratic} is often much easier than the original problem \eqref{prob:FP_mixed_matrix} thanks to the matrix ratio decoupling. The following application of FP for data clustering illustrates this point. For the data clustering problem, neither the concave-convex condition nor the convex-concave condition is
satisfied (because of the discrete constraints), but alternating
optimization can still be performed efficiently after decoupling the ratio. Furthermore, we remark that the matrix extension can be considered for the more general sum-of-functions-of-ratio problem \eqref{prob:FP_mixed} as discussed in \cite{Yannan_TSP24}.

\begin{figure*}[t]
  \centering
  \includegraphics[width=0.82\linewidth]{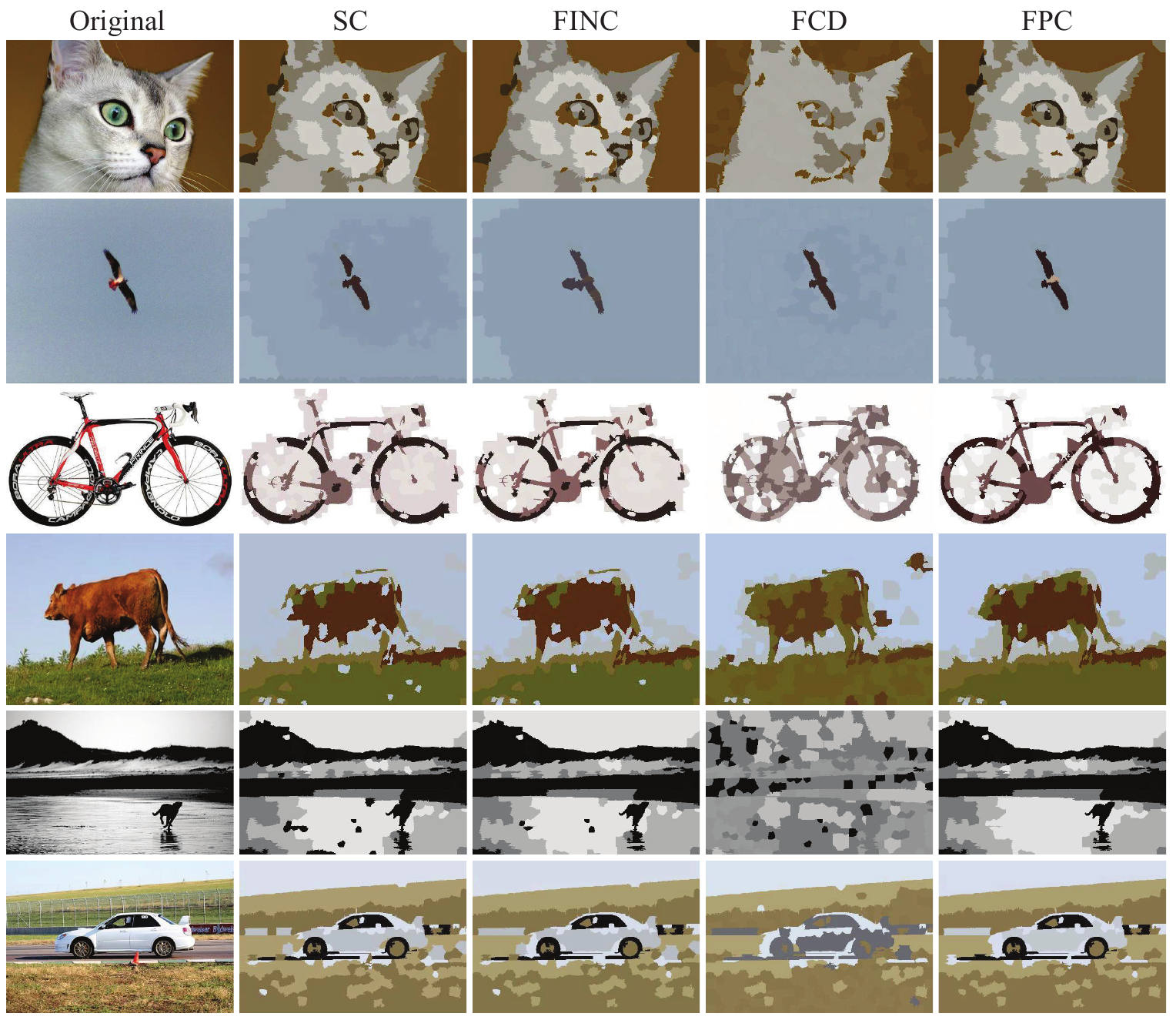}
  \caption{Image segmentation by the different algorithms based on the normalized-cut problem solving in Example \ref{example:ncut}. After decoupling the ratio in the normalized-cut problem by the quadratic transform, we convert the nonlinear discrete optimization to a weighted bipartite matching problem that can be readily solved. SC: spectral clustering; FINC: fast iterative normalized cut; FCD: fast coordinate descent; FPC: FP-based clustering.}
  \label{fig:seg}
\end{figure*}

\begin{example}[Normalized-Cut Problem for Data Clustering]
\label{example:ncut}
Suppose that there are $N$ data points in total. 
We use $i,j\in\{1,2,\ldots,N\}$ as indices for these data points. 
For a pair of data points $i$ and $j$, the similarity between them is
quantified as $0\le w_{ij}\le 1$. By symmetry, we have $w_{ij}=w_{ji}$. 
Visualizing in terms of a graph, the data points can be thought of as 
vertices, the similarities can be thought of as edges with weight $w_{ij}$
(or $w_{ji}$) assigned to the edge between vertex $i$ and vertex $j$.
Denote the set of vertices by $\mathcal V$, and the set of edges by
$\mathcal E$.  The data clustering problem can be expressed as a problem of partitioning a weighted undirected graph $G=(\mathcal V,\mathcal E)$, in which
the degree of each vertex $i$ is defined as $d_i=\sum^N_{j=1}w_{ij}$. 

The goal is to divide the $N$ data points into $K>1$ clusters. This is equivalent to partitioning $\mathcal V$ into $K$ disjoint subsets $\{\mathcal V_1,\mathcal V_2,\ldots,\mathcal V_K\}$, where $\bigcup^K_{k=1}\mathcal V_k=\mathcal V$ and $\mathcal V_k\cap\mathcal V_{k'}=\emptyset$ for any $k\ne k'$. The volume of each cluster $k$ is defined as $\mathrm{vol}(\mathcal V_k)=\sum_{i\in\mathcal V_k}d_i$. Intuitively, data clustering aims to group together those data points with high similarities between them.
But it is also important to regularize the cluster sizes, as otherwise the algorithm tends to put almost all data points in one giant cluster and leaves other clusters almost empty, leading to the cluster imbalance problem. The above goal can be accomplished by minimizing
\begin{equation}
\mathrm{ncut}(\mathcal V_1,\mathcal V_2,\ldots,\mathcal V_K)=\sum^K_{k=1}\frac{\sum_{i\in\mathcal V_k}\sum_{j\notin\mathcal V_k}w_{ij}}{\mathrm{vol}(\mathcal V_k)}.
\end{equation}
We can rewrite the problem by introducing 
indicator variable $x_{ik}\in\{0,1\}$, which equals to 1 if the data point is assigned to cluster $k$, and equals 0 otherwise. Let $\bx_k=[x_{1k},x_{2k},\ldots,x_{Nk}]^\top$ and $\bW=[w_{ij}]\in\mathbb R^{N\times N}$. We remark that $\bW$ is typically positive definite (e.g., when the similarities are generated by a Gaussian kernel). It can be shown that the normalized-cut minimization problem boils down to
\begin{equation}
%\label{prob:NCuts}
\begin{aligned}
\underset{ \bx }{\text{maximize}}& \quad \sum^K_{k=1}\frac{\bx_k^\top\bW\bx_k}{\bd^\top\bx_k}\\
\text{subject to}& \quad\, \bx_k\in\{0,1\}^N,\quad k=1,\ldots,K\\
&\quad\sum^K_{k=1}x_{ik} = 1,\quad i=1,\ldots,N,\\
%& \quad\, \bm{1}_m^\top\bx_i = 1,\quad i=1,\ldots,n,
\end{aligned}
\end{equation}
where $\bd=[d_1,d_2,\ldots,d_N]^\top$. The constraint $\sum^K_{k=1}x_{ik} = 1$ states that each data point $i$ can be assigned to only one cluster. The two constraints ensure that each data point must be assigned to a unique cluster. Because the problem is fractionally structured, it is natural to adopt an FP approach.

The authors of \cite{Chen_PAMI} suggest using the scalar quadratic transform to decouple the ratios in the above problem as
\begin{equation}
%\label{prob:NCuts}
\begin{aligned}
\underset{ \bx,\,  y}{\text{maximize}}& \quad \sum^K_{k=1}\Big(2y_k\sqrt{\bx_k^\top \bW\bx_k}-y^2_k{\bd^\top\bx_k}\Big)\\
\text{subject to}& \quad\, \bx_k\in\{0,1\}^N,\quad k=1,\ldots,K\\
&\quad\sum^K_{k=1}x_{ik} = 1,\quad i=1,\ldots,N\\
%& \quad\,\bm{1}_m^\top\bx_i = 1,\quad i=1,\ldots,n\\
&\quad\, y_k\in\mathbb R,\quad  k=1,\ldots,K.
\end{aligned}
\end{equation}
Further, \cite{Chen_PAMI} proposes to convexify the new optimization problem in $ \bx$ by the Cauchy-Schwarz inequality. The resulting alternating optimization between $ \bx$ and $  y$ guarantees monotonically decreasing convergence of $\mathrm{ncut}(\mathcal V_1,\mathcal V_2,\ldots,\mathcal V_K)$.

A better way to solve this problem is to apply the matrix quadratic transform. First, factorize the numerator as $\bx^\top_k\bW\bx_k = \bz_k\bz^\top_k$ where $\bz_k = \bx_k^\top\bW^{\frac12}\in\mathbb R^{1\times N}$ and $\bW^{\frac12}\in\mathbb S^{N\times N}_{+}$ is the symmetric square root of $\bW$. We then rewrite $(\bx_k^\top\bW\bx_k)/(\bd^\top\bx_k)$ as $\mathrm{Tr}(\bz_k^\top(\bd^\top\bx_k)^{-1}\bz_k)$. At first glance, it seems peculiar to rewrite the scalar ratio in a much more complicated matrix form, but this will pay off soon. By the matrix quadratic transform \eqref{prob:FP_mixed_matrix:quadratic}, we convert the original normalized-cut problem to
\begin{equation}
%\label{prob:NCuts}
\begin{aligned}
\underset{ \bx,\,\by}{\text{maximize}}& \quad \sum^K_{k=1}\mathrm{Tr}\big(2\by_k\bx_k^\top\bW^{\frac12}-\by_k{\bd^\top\bx_k}\by_k^\top\big)\\
\text{subject to}& \quad\, \bx_k\in\{0,1\}^N,\quad k = 1,\ldots,K\\
&\quad\sum^K_{k=1}x_{ik} = 1,\quad  i=1,\ldots,N\\
%& \quad\,\bm{1}_m^\top\bx_i = 1,\quad  i=1,\ldots,n\\
&\quad\, \by_k\in\mathbb R^N,\quad  k=1,\ldots,K.
\end{aligned}
\end{equation}
The key observation from \cite{shenNIPS} is that solving for $ \bx$ in the above new problem under fixed $ \by$ is a weighted bipartite matching problem that can be solved in closed form as
\begin{equation}
\label{eq:update X}
    x^\star_{ik}=
    \begin{cases}
        \;\;1\quad &\text{if }k = \mathop{\arg}\max\limits_{k'}\, \mu_{ik'}\vspace{0.5em}\\
        \;\;0\quad &\text{otherwise}
    \end{cases}, %\quad\text{for each}\; (i,k),
\end{equation}
where $\mu_{ik}$ is the $i$th component of $\bm\mu_{k} = 2\bW^{\frac12}\by_k-\by^\top_k\by_k\bd$.
This alternating optimization between $ \bx$ and $ \by$ is referred to as the \emph{fractional programming based clustering (FPC)} algorithm.

We illustrate the performance advantage of FPC in an image segmentation task. The benchmarks are the spectral clustering (SC) algorithm \cite{NCut}, the fast iterative normalized cut (FINC) algorithm \cite{Chen_PAMI}, and the fast coordinate descent (FCD) algorithm \cite{FCD}. We use the Gaussian kernel to generate the similarity matrix, i.e., $w_{ij} = \exp\left(-\|\bv_i-\bv_j\|_2^{2}\right)$, where $\bv_i$ and $\bv_j$ are the feature vectors of data points $i$ and $j$. As shown in Fig.~\ref{fig:seg}, clustering by FPC gives clearer boundaries of the objects than the other methods. 
\end{example}

\begin{example}[Pilot Signal Design for Channel Estimation]
\label{example:pilot}
Consider a wireless cellular network consisting of $L$ cells, with one BS and $K$ user terminals per cell. We use $i$ or $j$ to index each cell and its BS; the $k$th user in cell $i$ is indexed as $(i,k)$. Assume that every BS has $N$ antennas and every user terminal has a single antenna. Let $\bh_{ijk}\in\mathbb C^N$ be the channel from user $(j,k)$ to BS $i$, and let $\bH_{ij}=[\bh_{ij1},\bh_{ij2},\ldots,\bh_{ijK}]$, with each $\bh_{ijk}$ modeled as
$\bh_{ijk} = \bg_{ijk}\sqrt{\beta_{ijk}}$, 
where $\bg_{ijk}\in\mathbb C^{N}$ is the unknown small-scale fading coefficient with i.i.d. entries distributed as $\mathcal{CN}(0,1)$, and $\beta_{ijk}\ge0$ is the large-scale fading coefficient, assumed to be known. Each BS $i$ estimates its $\bH_{ii}$ based on the uplink pilot signals
from the users in the cell. Let $\bs_{ik}\in\mathbb C^\tau$ be a sequence of
pilot symbols transmitted from user $(i,k)$, and let $\bS_i=[\bs_{i1},\bs_{i2},\ldots,\bs_{iK}]$. Due to the limited pilot length, the pilot sequences across all the cells cannot all be orthogonal. This results in pilot contamination. We aim to design $\bS_1,\cdots,\bS_L$ based on the large-scale fading coefficients $\beta_{ijk}$ to minimize pilot contamination across the $L$ cells. 

The pilot signal received at BS $i$ is
\begin{equation}
\bV_i = \bH_{ii}\bS^\top_i + \sum\nolimits_{j\neq i}\bH_{ij}\bS^\top_j+\bZ_i,
\end{equation}
where $\bZ_i\in\mathbb C^{N\times \tau}$ is the additive noise with i.i.d. entries distributed as
$\mathcal{CN}(0,\sigma^2)$. Let $\hat \bh_{iik}$ be the minimum mean squared error (MMSE) estimate of $\bh_{iik}$ based on $\bV_i$. We aim to minimize the sum of mean squared errors (MSEs):
\begin{equation}
\label{weighted_MSE_min}
\underset{ {\bS}}{\text{minimize}}\quad\sum^L_{i=1}\sum^K_{k=1}\mathbb{E}\Big[\|\widehat{\bh}_{iik}-\bh_{iik}\|_2^2\mid \bV_i\Big].
\end{equation}
After a bit of algebra, the MSE minimization problem can be rewritten as
\begin{equation}
\begin{aligned}
\underset{ {\bS}}{\text{maximize}} &\quad
\sum^L_{i=1}\mathrm{Tr}\big(\bP_{ii}\bS^\hh_i\bD^{-1}_{i}\bS_i\bP_{ii}\big)\\
{\text{subject to}}&\quad\,  \|\bs_{ik}\|_2^2\le \rho,\quad\text{for any pair}\; (i,k),
\end{aligned}
\end{equation}
where $\bP_{ij}=\mathrm{diag}[\beta_{ij1},\beta_{ij2},\ldots,\beta_{ijK}]$, $\rho$ is the power constraint, and
$\bD_{i} = \sigma^2\bm I_\tau+\sum^L_{j=1}\bS_{j}\bP_{ij}\bS^\hh_{j}$. The above problem can be immediately addressed by the matrix quadratic transform, by treating $\sqrt{\bA_i}=\bS_i\bP_{ii}$ and $\bB_i=\bD_i$. The resulting pilot design is referred to as the \emph{fractional programming pilot (FPP)}.

\begin{figure}[t]
\centering
\centerline{\includegraphics[width=0.95\linewidth]{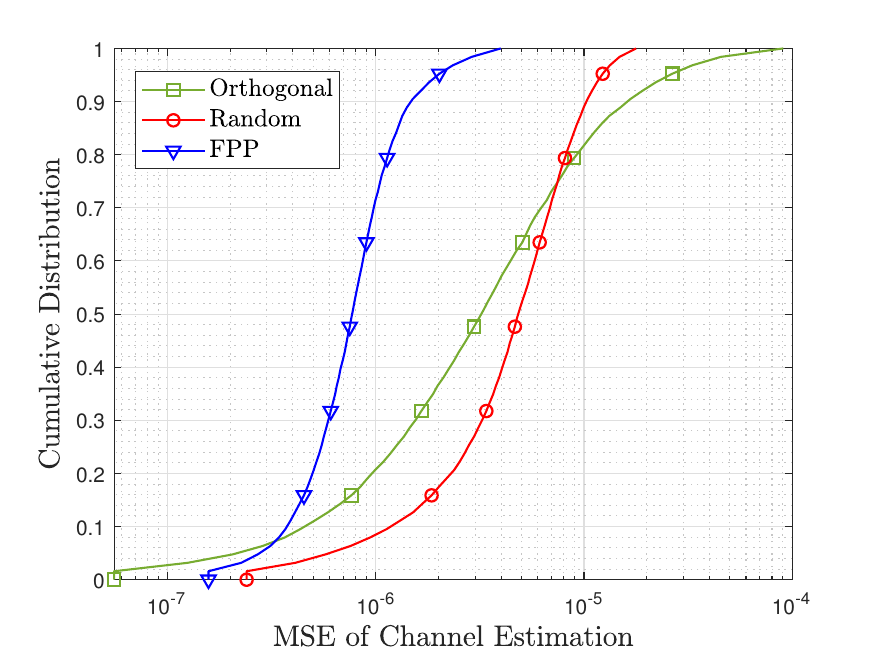}}
\caption{Cumulative distributions of the MSEs of channel estimation by the different pilot design methods in Example \ref{example:pilot}.}
\label{fig:pilot}
\end{figure}

We validate the performance of the FP method in a 7-cell wrapped-around
network. Each cell
comprises a 16-antenna BS and 9 single-antenna user
terminals uniformly distributed. The BS-to-BS distance is 1000 meters. Let
$\tau=10$ and let $\rho=1$. Assume that the
background noise is negligible and that
$\beta_{ijk}={\varphi_{ijk}}/{d_{ijk}^3}$ where $\varphi_{ijk}$ is an i.i.d.
log-normal random variable according to $\mathcal N(0,8^2)$ and $d_{ijk}$ is
the distance between user $(j,k)$ and BS $i$. Consider two baseline methods: 1) \emph{orthogonal method}, which fixes a set of 10 orthogonal pilots and allocates a
random subset of 9 pilots to users in each cell, and 2) \emph{random pilot method}, which generates the pilots randomly and independently according to the Gaussian distribution. The orthogonal method is used to initialize the FPP method. The simulation results are shown in Fig.~\ref{fig:pilot}. Observe that FFP achieves much smaller MSE overall than the benchmarks.
\end{example}

\section{Connections with Various Aspects of Optimization Theory}

\subsection{Connection with MM Method}
\label{subsec:MM}

We show that the quadratic transform \cite{shen2018fractional}, the Lagrangian dual transform \cite{shen2018fractional2}, and the AM-GM inequality transform \cite{ZhaoJun_JSAC} can all be interpreted as an MM method \cite{razaviyayn2013unified,sun2016majorization}.
A brief review of the MM method is as follows. For the primal problem
\begin{equation}
\label{MM f_o}
\begin{aligned}
    \underset{x\in\mathcal X}{\text{maximize}}& \quad f_o(x),
\end{aligned}
\end{equation}
where $f_o(x)$ is typically nonconcave, the MM method constructs a so-called \emph{surrogate function} $g(x|\hat x)$ conditioned on the parameter $\hat x$ that satisfies the following conditions:
\begin{equation}
\label{MM conditions}
g(x|\hat x)\le f_o(x)\;\;\forall x\in\mathcal X \;\;\text{and}\;\;
g(\hat x|\hat x)= f_o(\hat x).
\end{equation}
Then instead of optimizing $x$ directly in the primal problem, the MM method solves the new problem
\begin{equation}
\label{MM g}
\begin{aligned}
    \underset{x\in\mathcal X}{\text{maximize}}& \quad g(x|\hat x),
\end{aligned}
\end{equation}
where the condition variable $\hat x$ is updated to the previous solution $x$ iteratively. Optimizing $x$ for fixed $\hat x$ is referred to as the \emph{maximization} step, while updating $\hat x$ for the current $x$ is referred to as the \emph{minorization} step.

It turns out that the entire family of the quadratic transform methods can all be interpreted as MM methods. Let us take the unified quadratic transform as an example. In this case,
\begin{equation}
f_o(x)=\sum^{n^+}_{i=1}f^+_i\left(\frac{A^+_i(x)}{B^+_i(x)}\right)+\sum^{n^-}_{i=1}f^-_i\left(\frac{A^-_i(x)}{B^-_i(x)}\right).
\end{equation}
We now treat the optimal update of $y_i$ in \eqref{QT:opt y} as a function of $\hat x$, i.e., 
\begin{equation}
\mathcal Y_i(\hat x)=\sqrt{A^+_i(\hat x)}/B^+_i(\hat x),
\end{equation}
and replace each $y_i$ with $\mathcal Y_i(\hat x)$ in the new optimization objective in \eqref{prob:FP_mixed:quadratic}; likewise, we replace $\tilde y_i$ by 
$\widetilde{\mathcal Y}_i(\hat x)$ where 
\begin{equation}
\widetilde{\mathcal Y}_i(\hat x)=\sqrt{B^-_i(\hat x)}/A^-_i(\hat x).
\end{equation}
This gives a conditional function
\begin{align}
g(x|\hat x)& =\sum^{n^+}_{i=1}f_i^+\Big(2\mathcal Y_i(\hat x)\sqrt{A^+_i(x)}-\mathcal Y^2_i(\hat x)B^+_i(x)\Big) \nonumber \\ 
+ & \sum^{n^-}_{i=1}f_i^-\Big(\big[2\widetilde{\mathcal Y}_i(\hat x)\sqrt{B_i^{-}(x)}-\widetilde{\mathcal Y}^2_i(\hat x)A^-_i(x)\big]_+^{-1}\Big).
\end{align} 
It can be shown that the above $g(x|\hat x)$ satisfies the defining properties of MM in \eqref{MM conditions}, so it is a valid surrogate function. Thus, for the alternating optimization between $x$ and $  y$ by the unified quadratic transform, we can view the update of $x$ as the maximization step, and the update of $  y$ as the minorization step. 

Likewise, for the channel capacity maximization problem
\begin{equation}
f_o(x)=\sum^n_{i=1}w_i\log\left(1+\frac{A_i(x)}{B_i(x)}\right),
\end{equation}
the Lagrangian dual transform in \eqref{prob:log ratio:Lagrangian}
can be interpreted as constructing a surrogate function 
\begin{align}
g(x|\hat x)=\sum^n_{i=1}& w_i\Big(\log\big(1+\Gamma_i(\hat x)\big) \nonumber \\
& +\frac{(1+\Gamma_i(x))A_i(x)}{A_i(x)+B_i(x)}-\Gamma_i(\hat x)\Big),
\end{align}
where $\Gamma_i(\hat x) = \frac{A_i(\hat x)}{B_i(\hat x)}$, so it also belongs to the MM family.

Moreover, the AM-GM inequality transform from \cite{ZhaoJun_JSAC} for the sum-of-ratios min problem \eqref{prob:sum of ratios min} can be interpreted as an MM method as well. In this case, we have 
\begin{equation}
f_o(x)=-\sum^n_{i=1}\frac{A_i(x)}{B_i(x)}.
\end{equation}
As before, we treat the optimal update of the auxiliary variable $y_i$ in \eqref{Zhao-Qian-Yu opt y} as a function 
\begin{equation}
\mathcal Y_i(\hat x)=\frac{2}{{A_i(\hat x)B_i(\hat x)}},
\end{equation}
and then replace each $y_i$ with $\mathcal Y_i(\hat x)$ in the new objective in \eqref{prob:AM-GM} to obtain
\begin{equation}
g(x|\hat x)=-\sum^N_{n=1}w_i\left(\mathcal Y_i(\hat x)A^2_i(x)+\frac{1}{4\mathcal Y_i(\hat x)B^2_i(x)}\right).
\end{equation}
The fact that the above $g(x|\hat x)$ satisfies the defining properties of MM \eqref{MM conditions} follows from the AM-GM inequality:
\begin{equation}
\mathcal Y_i(\hat x)A^2_i(x)+\frac{1}{4\mathcal Y_i(\hat x)B^2_i(x)} \ge \frac{A_i(x)}{B_i(x)},
\end{equation}
where the equality holds if and only if we have $\mathcal Y_i(\hat x)A^2_i(x)=1/({4\mathcal Y_i(\hat x)B^2_i(x)})$, i.e., when $\mathcal Y_i(\hat x)$ equals $y_i$ in \eqref{Zhao-Qian-Yu opt y}.

The MM interpretation of these FP methods brings two benefits. First, it justifies the composition of the quadratic transform and other methods, e.g., the use of the Lagrangian dual transform in conjunction with the quadratic transform for power control as illustrated in Example \ref{example:power control}. This is because the composition is still an MM method. 
(In contrast, because Dinkelbach's method does not belong to the MM family, its combination with the other FP methods cannot be as easily justified.) 

Second, with the MM interpretation at hand, we can readily derive convergence conditions for the FP methods. The MM interpretation can further enable the convergence rate analysis as discussed later; intuitively, the convergence rate depends on how tight $g(x|\hat{x})$ approximates $f_o(x)$. We defer the detailed discussion to the later part of this article.

Since the quadratic transform can be interpreted as an MM method, it would be tempting to consider alternative MM methods by choosing other surrogate functions. But it is not always easy to determine the best surrogate function in practice. For instance, one may suggest using the first-order Taylor expansion, but the resulting linear surrogate function is less tight than the quadratic surrogate function in the quadratic transform; the tightness of various surrogate functions for MM is shown in Fig.~\ref{fig:MM} later in the article. Moreover, one may suggest using the second-order Taylor expansion to construct an alternative quadratic surrogate function, but its application is more limited than the quadratic transform, because it requires the ratio function to be continuously differentiable with a fixed Lipschitz constant; in contrast, the quadratic transform only requires the basic assumptions for FP, i.e., each numerator is nonnegative while each denominator is strictly positive. Further, it is possible to construct an even tighter surrogate function than the quadratic form for some specific cases, but this surrogate function may not be easy to optimize, not to mention potential generalizability issues.

\subsection{Connection with Gradient Projection: Accelerating the Quadratic Transform}
\label{subsec:gradient}

We now focus attention on the following type of matrix ratio:
\begin{align}
\label{matrix-ratio}
M_i=\big(\bA_i\bx_i\big)^\hh\Bigg(\sum^n_{j=1}\bB_{ij}\bx_{j}\bx_{j}^\hh\bB_{ij}^\hh\Bigg)^{-1}\big(\bA_i\bx_i\big),
\end{align}
where $\bx_j\in\mathbb C^m$, $\bA_i\in\mathbb C^{\ell\times m}$, and $\bB_{ij}\in\mathbb C^{\ell\times m}$, for $i,j=1,\ldots,n$. Typically, $\ell \ll m$. The above ratio term is of particular interest because many important metrics in practice can be written in this form, e.g., the CRB, the Fisher information, the SINR, and the normalized cut. Consider the sum-of-weighted-matrix-ratios problem involving a total of $n$ such matrix ratios:
\begin{equation}
\label{prob:MFP}
\begin{aligned}
  \underset{ \bx}{\text{maximize}} &\quad f_o( \bx):= \sum^n_{i=1} w_iM_i\\
  \text{subject to} & \quad \bx_i\in\mathcal X_i,\quad i=1,\ldots,n,
\end{aligned}
\end{equation}
where each $w_i>0$ is a strictly positive weight and each $\mathcal X_i$ is a nonempty convex constraint set on $\bx_i$. By the quadratic transform \cite{shen2018fractional}, the original optimization objective $f_o( \bx)$ can be recast to
\begin{equation}
f_q( \bx, \by)
    = \sum^n_{i=1}w_i\Bigg(2\Re\{\bx^\hh_i\bA^\hh_i\by_i\}-\sum^n_{j=1}\by^\hh_i\bB_{ij}\bx_j\bx^\hh_j\bB^\hh_{ij}\by_i\Bigg)
\end{equation}
with the auxiliary variable $\by_i\in\mathbb C^\ell$ introduced for each matrix ratio. When $ \bx$ is fixed, each $\by_i$ is optimally determined in closed form. When $ \by$ is fixed, each $\bx_i$ is optimally determined as
\begin{equation}
\label{QT:x}
    \bx^\star_i = \arg\min_{\bx_i\in\mathcal X_i}\big\|\bD^{\frac12}_i\big(\bx_i-w_i\bD_i^{-1}\bA^\hh_i\by_i\big)\big\|_2,
\end{equation}
where $\bD_i = \sum^n_{j=1}w_j\bB_{ji}^\hh\by_j\by^\hh_j\bB_{ji}$.
A practical issue here is that the matrix inverse in \eqref{QT:x} can be quite costly when $\bD_i$ is a large matrix. 

\RestyleAlgo{ruled}
%\SetAlCapSkip{0.5em}
{\setstretch{0.6}
\begin{algorithm}[t]
initialize $ \bx$ to a feasible value\;
\Repeat{the value of $f_o( \bx)$ converges}{
update each $\by_i=\big(\sum^n_{j=1}\bB_{ij}\bx_j\bx^\hh_j\bB^\hh_{ij}\big)^{-1}\big(\bA_i\bx_i\big)$\;
    \Switch{the choice of transform}{
        \Case{basic quadratic transform}{update each $\bx_i$ according to \eqref{QT:x}\;}
        \Case{nonhomogeneous quadratic transform}{update each $\bx_i$ according to \eqref{GQT:x}\;}
        \Case{extrapolated quadratic transform}{update each $\bm{\nu}_i$ according to \eqref{extrapolate}\;
        update each $\bx_i$ according to \eqref{extrapolate:x}\;}
    }
}
\caption{Different approaches to problem \eqref{prob:MFP}.}
\label{algorithm:EQT}
\end{algorithm}
}

The so-called \emph{nonhomogeneous quadratic transform} \cite{ZP_MM+} aims to eliminate the matrix inverse operation. The main idea is to construct a lower bound on 
$f_q( \bx, \by)$ as
\begin{multline}  
\label{ft}
f_t( \bx, \by, \bz) = \sum^n_{i=1} \Big( 2\Re\big\{w_i\bx^\hh_i\bA^\hh_i\by_i+\bx^\hh_i(\lambda_i\bI-\bD_i)\bz_i\big\}\\
+\bz^\hh_i(\bD_i-\lambda_i\bI)\bz_i-\lambda_i\bx^\hh_i\bx_i\Big),
\end{multline}
where $\lambda_i\ge\lambda_{\max}(\bD_i)$, e.g., $\lambda_i=\|\bD_i\|_\mathrm{F}$. We have $f_q( \bx, \by)\ge f_t( \bx, \by, \bz)$ in general, where the equality holds if $ \bz= \bx$. As a consequence, the problem of maximizing $f_q( \bx, \by)$ over $\bx$ and $\by$ is equivalent to the problem of maximizing $f_t( \bx, \by, \bz)$ over $\bx$, $\by$ and $\bz$. We then consider optimizing $ \bx$, $ \by$, and $ \bz$ iteratively in $f_t( \bx, \by, \bz)$. When $ \by$ and $ \bx$ are both held fixed, the optimal update of $ \bz$ is $\bz_i^\star = \bx_i$. When $ \bz$ and $ \bx$ are both fixed, the optimal update of each $\by_i$ is 
\begin{equation}
\label{GQT:y}
\by_i=\left(\sum^n_{j=1}\bB_{ij}\bx_j\bx^\hh_j\bB^\hh_{ij}\right)^{-1}\big(\bA_i\bx_i\big).
\end{equation}
Next, when $ \by$ and $ \bz$ are both held fixed, the optimal $\bx_i$ is given by
\begin{equation}
\label{GQT:x}
    \bx_i^\star = \arg\min_{\bx_i\in\mathcal X_i}\big\|\lambda_i\bx_i-w_i\bA^\hh_i\by_i-\big(\lambda_i\bI-\bD_i\big)\bz_i\big\|_2.
\end{equation}
Observe that the computation of the matrix inverse of the $m \times m$ matrix $\bD_i$ is no longer required. Instead, the only matrix inverse needed is in the update of $\by$ in \eqref{GQT:y}. Note that the matrix inverse in \eqref{GQT:y} is typically of a much smaller dimension, i.e., it is $\ell \times \ell$ instead of $m \times m$ as in \eqref{QT:x}.

Interestingly, the nonhomogeneous quadratic transform has an intimate connection with gradient projection. We use the superscript $k\in\{1,2,\ldots\}$ to index the iteration, and let $( \bx, \by, \bz)$ be cyclically updated as
$ \bx^{(0)}\rightarrow\cdots\rightarrow  \bz^{(k)} \rightarrow  \by^{(k)} \rightarrow  \bx^{(k)} \rightarrow  \bz^{(k+1)} \rightarrow \cdots$.
We can then rewrite the optimal update of $\bx_i$ in \eqref{GQT:x} as the Euclidean projection onto the constraint set $\mathcal X_i$:
\begin{equation}
\label{sphere projection}
    \bx^{k}_i 
    = \mathcal{P}_{\mathcal X_i}\bigg(\bz_i^{(k)} + \frac{1}{\lambda_i}\Big(w_i\bA^\hh_i\by_i^{(k)}-\bD_i\bz^{(k)}_i\Big)\bigg),
\end{equation}
which can be further recognized as gradient projection \cite{Shen_JSAC24}:
\begin{align}
    \bx^{(k)}_i 
    &= \mathcal{P}_{\mathcal X_i}\bigg(\bx^{(k-1)}_i + \frac{1}{\lambda_i}\cdot\frac{\partial f_o( \bx^{(k-1)})}{\partial \bx_i}\bigg).
\end{align}

\begin{figure}[t]
\centering
%\subfigure[]
\includegraphics[width=0.85\linewidth]{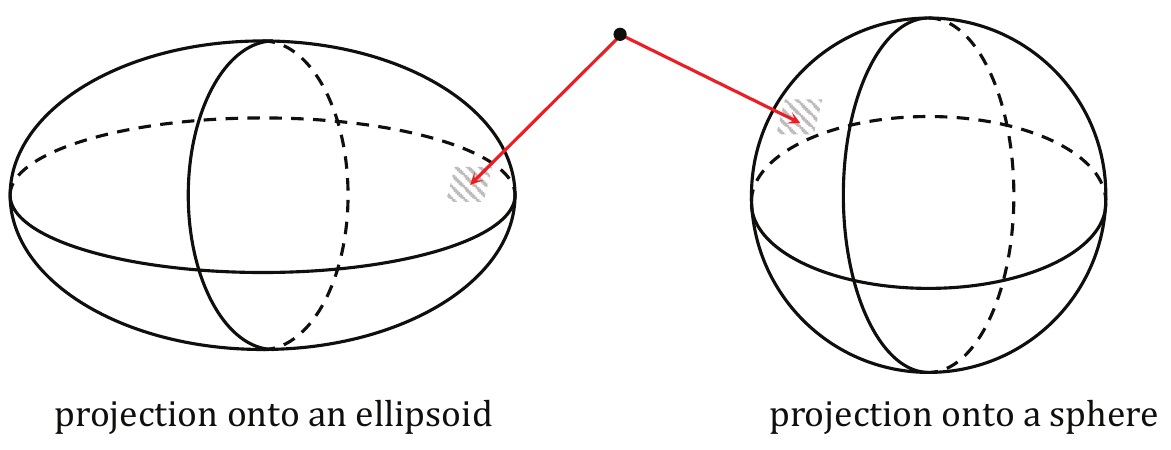}
\caption{The basic quadratic transform can be interpreted as the projection onto an ellipsoid in \eqref{QT:x}, while the nonhomogeneous quadratic transform can be interpreted as the projection onto a sphere in \eqref{GQT:x}. The nonhomogeneous quadratic transform eliminates the matrix inversion when solving the matrix-ratio problem \eqref{prob:MFP}.}
\vspace{1em}
\label{fig:projection}
\end{figure}

We now compare the basic quadratic transform and the nonhomogeneous quadratic transform from a geometric perspective. As shown in Fig.~\ref{fig:projection}, the update of $\bx_i$ in \eqref{QT:x} by the basic quadratic transform can be interpreted as the projection onto the surface of an ellipsoid, while the update of $\bx_i$ in \eqref{GQT:x} by the nonhomogeneous quadratic transform can be interpreted as the projection onto the surface of a sphere. Projection onto an ellipsoid is in general much more costly to do than projection onto a sphere in a high-dimensional space.

The fact that the nonhomogeneous quadratic transform amounts to a gradient projection method motivates us to investigate the possibility of incorporating Nesterov's extrapolation scheme \cite{Nesterov_book} into FP in order to accelerate its convergence. Specifically, following the \emph{heavy-ball} intuition, we extrapolate each $\bx_i$ along the direction of the difference between the preceding two iterates before the gradient projection, i.e.,
\begin{align}
    \bm{\nu}^{(k-1)}_i &= \bx^{(k-1)}_i+\eta_{k-1}(\bx^{(k-1)}_i-\bx^{(k-2)}_i),
    \label{extrapolate}\\
    \bx^{(k)}_i &= \mathcal{P}_{\mathcal X}\bigg(\bm{\nu}_i^{(k-1)} + \frac{1}{\lambda_i}\cdot\frac{\partial f_o( {\bm{\nu}}^{(k-1)})}{\partial \bx_i}\bigg),
    \label{extrapolate:x}
\end{align}
where the extrapolation stepsize 
$\eta_k = \max\big\{\frac{k-2}{k+1},0\big\}$ and the starting point $\bx^{(-1)}=\bx^{(0)}$, as in \cite{Nesterov_book}. We refer to this algorithm as the \emph{extrapolated quadratic transform}. Moreover, a recent work \cite{Hien2023} suggests that the extrapolation scheme can be incorporated into the MM method, so it is possible to accelerate the quadratic transform directly without using the nonhomogeneous quadratic transform. Algorithm \ref{algorithm:EQT} summarizes the basic, the nonhomogeneous, and the extrapolated quadratic transforms.

\subsection{Connection with WMMSE Algorithm}
\label{subsec:WMMSE}

We now discuss the connections of FP to specific optimization techniques in wireless communications. The WMMSE algorithm \cite{chris,luo_wmmse} is a well-known algorithm for computing the optimal beamformers in a multi-cell MIMO transmission scenario. This section aims to show that the WMMSE algorithm is a special case of the FP method, and moreover such connection enables further improvement of the WMMSE algorithm.

Consider a wireless cellular network with $L$ cells and assume that there are $K$ downlink users in each cell. Assume also that each BS has $M$ transmit antennas and each user has $N$ receive antennas. The $k$th user in the $i$th cell is indexed as $(i,k)$. Denote by $\bH_{ik,j}\in\mathbb C^{N\times M}$ the channel from BS $j$ to user $(i,k)$, $\sigma^2$ the background noise power, and $\bV_{ik}\in\mathbb C^{M\times d}$ the transmit beamformer of BS $i$ intended for user $(i,k)$, where $d$ is the number of data streams. Let $\bB_{ik} = \sigma^2\bI+\sum_{(j,q)\ne(i,k)}\bH_{ik,j}\bV_{jq}\bV^\hh_{jq}\bH^\hh_{ik,j}$ and let $\sqrt{\bA_{ik}}=\bH_{ik,i}\bV_{ik}$.
The weighted-sum-rate maximization problem aims to maximize the objective of
\begin{align}
    f_o( \bV) =  \sum_{(i,k)}w_{ik}\log\det\big(\bI+\sqrt{\bA_{ik}}^\hh\bB_{ik}^{-1}\sqrt{\bA_{ik}}\big),
\end{align}
where the weight $w_{ik}>0$ reflects the priority of user $(i,k)$. While the previous works \cite{chris,luo_wmmse} obtain the WMMSE algorithm from the relationship between SINR and MMSE, we rederive WMMSE by purely using FP. First, by the matrix extension of the Lagrangian dual transform \cite{shen2019optimization}, the original objective function can be converted to
\begin{align}
&f_r(\bV, {\bm\Gamma}) =  \sum_{(i,k)}\Big(w_{ik}\log\det(\bI+\bm\Gamma_{ik})-w_{ik}\mathrm{Tr}\big(\bm\Gamma_{ik}\big)\notag\\
&\quad +w_{ik}\mathrm{Tr}\big((\bI+\bm\Gamma_{ik})\sqrt{\bA_{ik}}^\hh(\bA_{ik}+\bB_{ik})^{-1}\sqrt{\bA_{ik}}\big)\Big).
\end{align}
When $ \bV$ is fixed, each $\bm\Gamma_{ik}$ is optimally updated as $\bm\Gamma_{ik}^\star=\sqrt{\bA_{ik}}^\hh\bB_{ik}^{-1}\sqrt{\bA_{ik}}$. It remains to optimize $ \bV$ for fixed $ {\bm\Gamma}$. Maximizing $f_r( \bV, {\bm\Gamma})$ over $ \bV$ for fixed $ {\bm\Gamma}$ can be recognized as a sum-of-matrix-ratios problem. The quadratic transform can be performed in different ways: one particular way leads to WMMSE \cite{chris,luo_wmmse}, while another leads to the FPLinQ algorithm in \cite{shen2018fractional2,shen2019optimization}. Thus, WMMSE can be interpreted as a particular case of the quadratic transform method.

We now show how to obtain the WMMSE algorithm from the quadratic transform. If we treat $f_r( \bV, {\bm\Gamma})$ as a weighted sum of the following matrix ratios
$$
\underbrace{\big(\bA_{ik}\big)}_{\text{numerator}}{\underbrace{(\bA_{ik}+\bB_{ik})}_{\text{denominator}}}^{-1}
$$
and use the quadratic transform to decouple each matrix ratio, then  $f_r( \bV, {\bm\Gamma})$ can be further recast as
\begin{align}
&f^{(\mathrm{I})}_q( \bV, {\bm\Gamma}, \bY) =\sum_{(i,k)}\Big(w_{ik}\log\det(\bI+\bm\Gamma_{ik})-w_{ik}\mathrm{Tr}\big(\bm\Gamma_{ik}\big)\,+\notag\\
&\;w_{ik}\mathrm{Tr}\big((\bI+\bm\Gamma_{ik})(\sqrt{\bA_{ik}}^\hh\bY_{ik} + \bY_{ik}^\hh\sqrt{\bA_{ik}} -\bY^\hh_{ik}\bB'_{ik}\bY_{ik})\big)\Big)
\end{align}
with $\bB'_{ik}=\bA_{ik}+\bB_{ik}$. Optimizing $( \bV, {\bm\Gamma}, \bY)$ iteratively in the above new objective function gives rise exactly to the WMMSE algorithm \cite{chris,luo_wmmse}.

\begin{figure}[t]
\centerline{\includegraphics[width=0.9\linewidth]{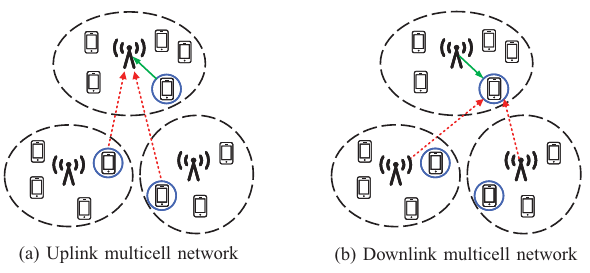}}
\caption{Interference pattern depends on the user scheduling in the neighboring cells in the uplink, but not so in the downlink. Here, the solid lines represent the desired signal; the dashed lines represent the interfering signal; the
scheduled user terminal in each cell is circled. The quadratic transform \cite{shen2018fractional2} is more powerful than WMMSE \cite{chris,luo_wmmse}, when the problem includes the discrete optimization of uplink user scheduling.}
\label{fig:uplink}
\end{figure}

There are also other ways of using the quadratic transform to decouple the matrix ratios. For instance, we could have alternatively regarded $f_r( \bV, {\bm\Gamma})$ as a sum of following matrix ratios:
$$
\underbrace{\big(w_{ik}(\bI+\bm\Gamma_{ik})\bA_{ik}\big)}_{\text{numerator}}\;{\underbrace{(\bA_{ik}+\bB_{ik})}_{\text{denominator}}}^{-1}.
$$
Differing from the preceding matrix ratio of the WMMSE case, the above matrix ratio includes the weight $w_{ik}$ and the auxiliary variable $\bI+\bm\Gamma_{ik}$ in the numerator. After decoupling the above matrix ratio by the quadratic transform, we recast $f_r( \bV, {\bm\Gamma})$ into a different objective function:
\begin{align}
&f^{(\mathrm{II})}_q( \bV, {\bm\Gamma}, \bY) =\sum_{(i,k)}\Big(w_{ik}\log\det(\bI+\bm\Gamma_{ik})-w_{ik}\mathrm{Tr}\big(\bm\Gamma_{ik}\big) \nonumber \\
&\quad +\sqrt{\bA'_{ik}}^\hh\bY_{ik} + \bY_{ik}^\hh\sqrt{\bA'_{ik}} -\bY^\hh_{ik}(\bA_{ik}+\bB_{ik})\bY_{ik}\Big)
\end{align}
with $\sqrt{\bA'_{ik}}=\sqrt{w_i}(\bI+\bm\Gamma_i)^{\frac12}\sqrt{\bA_{ik}}$. Again, we consider optimizing $( \bV, {\bm\Gamma}, \bY)$ in an iterative fashion. This gives rise to a beamforming algorithm called FPLinQ in \cite{shen2018fractional2,shen2019optimization}. We remark that the above results can be easily extended to more complicated networks. For instance, for the cell-free network \cite{Bjornson_cellfree21} wherein each downlink user is served by multiple BSs, the matrix ratio becomes 
$\big(\sum_{i\in\mathcal N_k}\sqrt{\bA_{ik}}\big)^\hh\bB_{k}^{-1}\big(\sum_{i\in\mathcal N_k}\sqrt{\bA_{ik}}\big)$,
where $\mathcal N_k$ is the set of BSs assigned to user $k$, $\sqrt{\bA_{ik}}$ is the precoded signal sent from BS $i$ intended for user $k$, and $\bB_k$ is the interference-plus-noise covariance matrix at user $k$. The resulting FP problem is mathematically the same as the cellular network discussed above.

Regardless of how the matrix ratio is decoupled by the quadratic transform, the FP method always introduces some auxiliary variables and then optimizes them along with the beamforming variable iteratively. This approach enjoys two advantages: (i) each iterative update can be done in closed form; (ii) it guarantees convergence to a stationary point of the beamforming problem (as shown earlier due to the connection to the MM theory). But are these distinct FP methods (e.g., WMMSE and FPLinQ) equally good? Does it matter in algorithm design which way to decouple the ratio?

\begin{figure*}
    \centering    \includegraphics[width=0.9\linewidth]{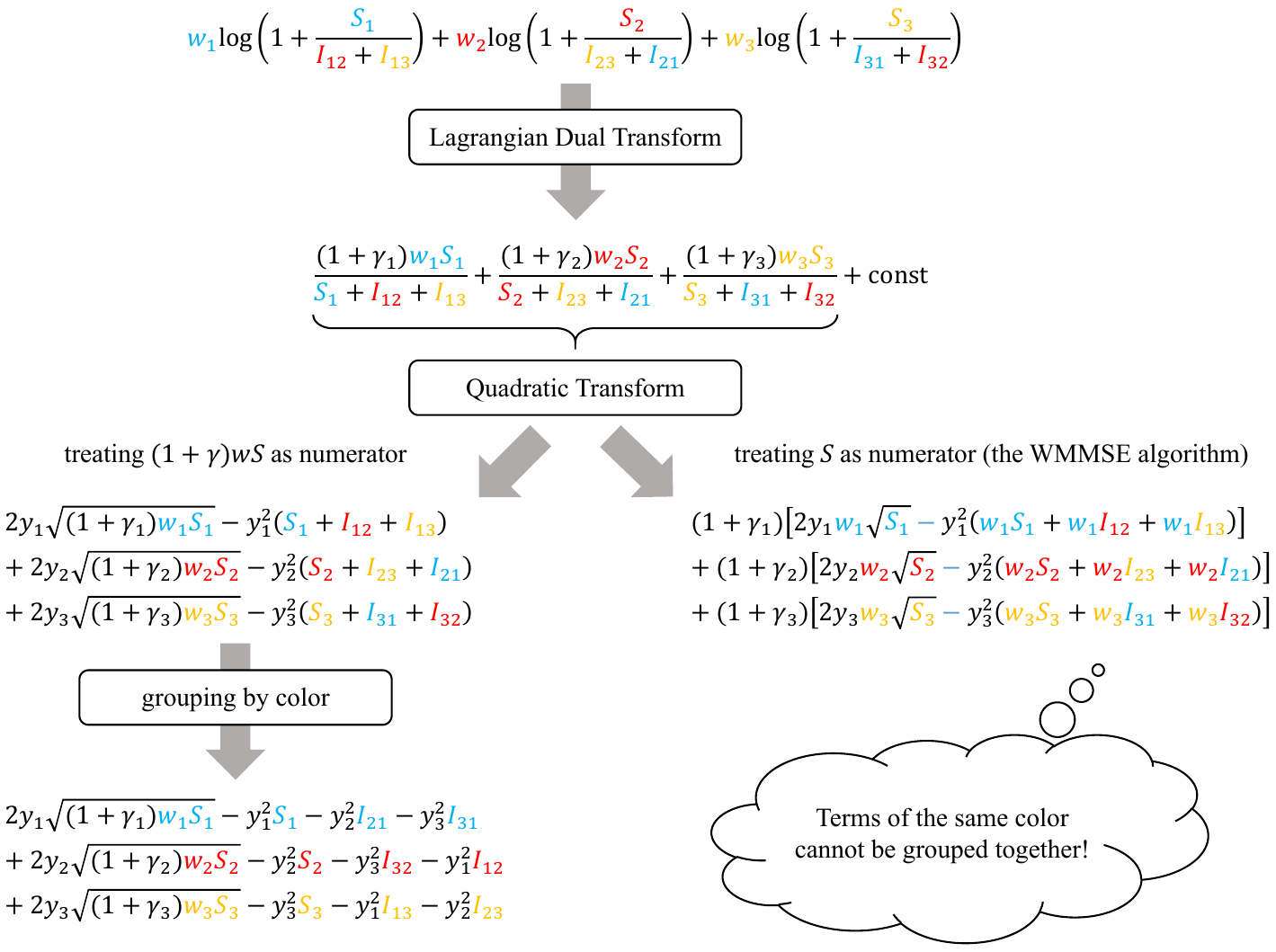}
    \caption{Consider uplink scheduling for three SISO cells. Let $w_i$ (the rate weight of the scheduled user), $S_i$ (the received signal power), and $I_{ij}$ (the interference from cell $j$ to cell $i$) be colored by their respective cell index $i$. After the Lagrangian dual transform, the optimization objective is transformed into a sum-of-weighted-ratios objective plus a constant. In the next step, if we apply the quadratic transform as in FPLinQ \cite{shen2018fractional2}, then the terms in the new problem can be grouped according to their colors, so that the scheduling problem becomes a weighted bipartite matching problem. In contrast, if we follow the WMMSE algorithm \cite{chris,luo_wmmse}, then the variables with distinct colors are coupled, making the discrete optimization problem difficult to tackle.}
    \label{fig:WMMSE}
\end{figure*}

The following example sheds light on the above questions. Let us shift attention to the uplink network. To ease notation, consider the single-input single-output (SISO) case, but the following discussion can be generalized to the MIMO case as in \cite{shen2018fractional2,shen2019optimization}. We incorporate uplink scheduling into the problem formulation, i.e., in each time-slot, we need to select one user in each cell for uplink transmission with an aim of maximizing a utility function of long-term average rates across all the users. It is worth noting that the uplink scheduling problem is more challenging than the downlink, as illustrated in Fig.~\ref{fig:uplink}. We can carry out the quadratic transform in different ways. If we decouple the ratios as in $f^{(\mathrm{I})}_q$, then the ratio decoupling does not make the scheduling problem much easier since the discrete scheduling variables are still coupled together in the new problem. But if we instead decouple the ratios as in $f^{(\mathrm{II})}_q$, then the discrete variables can be decoupled after the ratio decoupling, so the scheduling problem can be efficiently solved by the standard weighted bipartite matching method. Fig.~\ref{fig:WMMSE} gives a detailed exposition of the advantage of FPLinQ over WMMSE for the uplink scheduling problem.

\subsection{Connection with Fixed-Point Iteration}
\label{sec:fixed point}

We now revisit the earlier example of power control for interfering links, with an aim to connect the FP method with the fixed-point iteration based power control method \cite{wang_vt,hayssam,yates}. We use $f_o(  p)$ to denote the weighted-sum-rate objective of the power control problem, i.e.,
\begin{equation}
f_o(  p)=\sum^K_{i=1}w_i\log\Bigg(1+\frac{|h_{ii}|^2p_i}{\sum_{j\ne i}|h_{ij}|^2p_j+\sigma^2}\Bigg).
\end{equation}
For ease of discussion, we ignore the power constraint. Fixed-point iteration is an approach to the power control problem in the literature \cite{wang_vt,hayssam,yates} based on rewriting the first-order equation $\partial f_o(  p)/\partial p_i=0$ as $p_i=G_i(  p)$ whereby $p_i$ is isolated on the left-hand side and is iteratively updated by the previous $p_i$ on the other side, i.e.,  
\begin{equation}
p_i^{(t+1)}=G_i(  p^{(t)}),
\end{equation}
where the superscript $t$ or $t+1$ is the iteration index. Note that $\partial f_o(  p)/\partial p_i=0$ can be rewritten as $p_i=G_i(  p)$ in different ways, each leading to a different fixed-point iteration. If the iterative update of $p_i$ converges for every $i$, then the first-order condition must hold after convergence and thus we end up with a stationary-point solution. 

However, it is not an easy task to prove the convergence of a fixed-point iteration. A variety of fixed-point iteration methods have been proposed in \cite{wang_vt,hayssam,yates} in pursuit of convergence. For instance, the authors of \cite{hayssam} suggest rewriting $\partial f_o(  p)/\partial p_i=0$ as
\begin{equation}
%\label{Hayssam:fixed point}
p_{i}=\bigg(\frac{w_{i}\Gamma_i}{1+\Gamma_i}\bigg)
\Bigg(\sum^K_{j=1,j\neq i}
\frac{w_{j}\Gamma^2_i|h_{ji}|^2}
{(1+\Gamma_i)|h_{jj}|^2p_{j}}\Bigg)^{-1},
\end{equation}
with 
\begin{equation}
\Gamma_i=\frac{|h_{ii}|^2p_i}{\sum_{j\ne i}|h_{ij}|^2p_j+\sigma^2}.
\end{equation}
It is shown in \cite{hayssam} that the above fixed-point iteration can guarantee convergence provided that the initial value of every $\Gamma_i(  p^{(0)})$ is sufficiently high, by exploiting the \emph{standard function properties} \cite{yates}. However, finding a fixed-point iteration with provable convergence in general remains difficult. 

Through reverse engineering as shown in \cite{shen2018fractional}, the (closed-form) FP method for the power control problem as discussed in the earlier example can be interpreted as a fixed-point iteration with $\partial f_o(  p)/\partial p_i=0$ rewritten as
\begin{equation}
p_i = 
\bigg(\frac{w_i^2\Gamma_i^2}
{p_i}\bigg)
\Bigg(\sum^K_{j=1} \frac{w_j\Gamma^2_i|h_{ji}|^2}
{(1+\Gamma_i)|h_{jj}|^2p_j}\Bigg)^{-2}.
\end{equation}
The convergence of this fixed-point iteration now follows directly from the convergence of the FP method---which is verified by the MM theory as discussed earlier in the article.

\section{Convergence Analysis}
\label{sec:analysis}

This section presents two main results on the convergence of the FP algorithms. First, we show that the iterative optimization by the quadratic transform has provable convergence to a stationary-point solution of the original problem, under a weaker condition than that for the block coordinate descent (BCD) method. Second, focusing on the sum-of-weighted-matrix-ratios problem \eqref{prob:MFP}, we examine the rate of convergence for the different quadratic transform methods as summarized in Algorithm \ref{algorithm:EQT}.

The analysis relies on the MM theory
\cite{razaviyayn2013unified,sun2016majorization}. First of all, it is easy
to verify a composition property for the surrogate function in \eqref{MM
conditions}: if $g(x|\hat x)$ is a surrogate function of $f_o(x)$ while
$h(x|{\hat x})$ is a surrogate function of $g(x|x)$, then $h(x|{\hat x})$
must be a surrogate function of $f_o(x)$. We have already shown that the
quadratic transform (including all its extensions) can be interpreted as
MM. The Lagrangian dual transform in \eqref{prob:log ratio:Lagrangian} can
also be interpreted as MM. Thus, their composition, as considered in the
example of power control for interfering links, also amounts to
an MM method. Likewise, because the nonhomogenous bound in
\cite{sun2016majorization} in essence is about constructing a surrogate function,
its use in conjunction with the quadratic transform, namely the
nonhomogeneous quadratic transform, boils
down to an MM method too. Thus, by the MM theory, all these FP methods
guarantee that the value of the maximization (resp. minimization) objective
function of the original FP problem is nondecreasing (resp. nonincreasing)
after each iteration. Further, they converge to a stationary point of the
original FP problem under certain mild conditions as stated in
\cite{razaviyayn2013unified,sun2016majorization}. 

For instance, the alternating optimization between $x$ and $y$ in \eqref{prob:quadratic transform} guarantees convergence to a stationary point of the original problem \eqref{prob:sum of ratios} provided that each $A_i(x)$ is differentiable and concave, each $B_i(x)$ is differentiable and convex, and $\mathcal X$ is a convex set. Without using the MM interpretation, the alternating optimization can only be viewed as the BCD method, but the convergence condition for BCD is stronger, requiring each iterate to be uniquely solvable \cite{Bertsekas_book}. As such, $A_i(x)$ being concave and $B_i(x)$ being convex in \eqref{prob:sum of ratios} are no longer enough for convergence; it now requires strict concavity and strict convexity.

We further analyze how fast the different quadratic transform methods converge for the problem instance \eqref{prob:MFP} by considering the convergence of the function value versus the number of iterates. Consider the basic quadratic transform, the nonhomogeneous quadratic transform, and the extrapolated quadratic transform. Assume that the Hessian of the original objective function $f_o( \bx)$ is $L$-Lipschitz continuous. To make the analysis tractable, we further require the constraint set to be a small neighborhood of a local maximum $ \bx^*$, so that the distance between $ \bx^*$ and the $k$th-iteration solution $ \bx^{(k)}$ is bounded, i.e.,
$\| \bx- \bx^*\|_2\le R$ for all $k$. We then analyze the local convergence behavior assuming that the radius $R$ is sufficiently small. Recall that both the basic quadratic transform and nonhomogeneous quadratic transform are equivalent to constructing a surrogate function for $f_o( \bx)$. Let $g( \bx|\hat{ \bx})$ be the surrogate function due to either the basic quadratic transform or nonhomogeneous quadratic transform; define the gap between the original optimization objective $f_o( \bx)$ and the surrogate function $g( \bx|\hat{ \bx})$ to be $\delta( \bx|\hat{ \bx}) = f_o( \bx) -g( \bx|\hat{ \bx})$.
Moreover, let the maximum eigenvalue of the Hessian of $\delta( \bx|\hat{ \bx})$ be $\Lambda$. According to \cite{Shen_JSAC24}, the basic quadratic transform and nonhomogeneous quadratic transform have the following convergence rate:
\begin{equation}
f_o( \bx^*)-f_o( \bx^{(k)}) \le \frac{2\Lambda R^2+2LR^3/3}{k+3},\quad\text{for}\;k\ge2,
\end{equation}
where the parameter $\Lambda\ge0$ differs for the two transforms. It can be shown that $\Lambda$ of the basic quadratic transform is smaller than that of the nonhomogeneous quadratic transform, so the former converges faster as indicated by the above convergence rate analysis.

\begin{figure}[t]
\centering
%\subfloat[]
\includegraphics[width=1.0\linewidth]{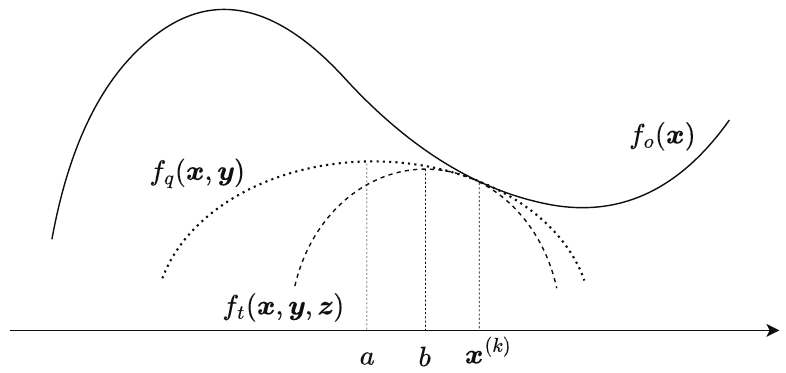}
\caption{The basic quadratic transform approximates $f_o( \bx)$ as $f_q( \bx, \by)$ while the nonhomogeneous quadratic transform approximates $f_o( \bx)$ as $f_t( \bx, \by, \bz)$. By the MM procedure, for the current solution $ \bx^{(k)}$, the basic quadratic transform updates it to $a$, while the nonhomogeneous quadratic transform updates it to $b$. The basic quadratic transform converges faster in iterations, because its approximation is tighter.}
\label{fig:MM}
\end{figure}

We provide an intuitive reason for the preceding conclusion. The parameter $\Lambda$ is proportional to the gap between the surrogate function and the original objective function. More precisely, $\Lambda$ reflects how well $g( \bx|\hat{ \bx})$ approximates $f_o( \bx)$ in terms of the second-order profile. In principle, a smaller $\Lambda$ leads to a tighter approximation. Recall that the basic quadratic transform uses the quadratic transform to construct a lower bound on $f_o( \bx)$, while the nonhomogeneous quadratic transform uses the nonhomogeneous bound \cite{sun2016majorization} to further bound the above lower bound from below, i.e., 
\begin{equation}
f_o( \bx)\ge f_q( \bx|\hat{ \bx})\ge f_t( \bx|\hat{ \bx}),
\end{equation}
where $f_q( \bx|\hat{ \bx})$ is the new objective function used in the basic quadratic transform, while $f_t( \bx|\hat{ \bx})$ is the new objective function used in the nonhomogeneous quadratic transform. Although both $f_q( \bx|\hat{ \bx})$ and $f_t( \bx|\hat{ \bx})$ are surrogate functions of $f_o( \bx)$, the former is closer to $f_o( \bx)$ and gives a tighter approximation as illustrated in Fig.~\ref{fig:MM}, so the basic quadratic transform converges faster. We emphasize that the convergence rate considered here is in terms of the number of iterations. In practice, the nonhomogeneous quadratic transform can run faster than the basic quadratic transform, because it avoids a large matrix inversion in each iteration, even though the nonhomogeneous quadratic transform may require more iterations to converge.

Since $\Lambda$ measures the gap between the surrogate function and $f_o( \bx)$, the best scenario one can hope for is $\Lambda=0$. This can be achieved by the \emph{cubically regularized Newton's method} due to Nesterov \cite{Nesterov_book}. As shown in Theorem 4.1.4 in \cite{Nesterov_book}, the resulting convergence rate is
\begin{equation}
\begin{aligned}
f_o( \bx^*)-f_o( \bx^{(k)}) &\le \frac{LR^3}{2(1+k/3)^2},\quad\text{for}\;k\ge2.
\end{aligned} 
\end{equation}
However, the cubically regularized Newton's method requires computing the inverse of the Hessian matrix and thus is costly in practice. 
We now show that the extrapolated quadratic transform can yield almost equally good performance. Recall that the extrapolated quadratic transform accelerates the nonhomogeneous quadratic transform by incorporating Nesterov's extrapolation scheme, so the convergence rate analysis for the momentum method in \cite{Nesterov_book} carries over to the extrapolated quadratic transform method. Suppose that the gradient of $f_o( \bx)$ is $C$-Lipschitz continuous. Then under certain conditions \cite{Nesterov_book} the extrapolated quadratic transform yields
\begin{equation}
    f( \bx^*) - f( \bx^{(k)}) \le \frac{2C\|\bx^*-\bx^{(0)}\|^2_2}{(k+1)^2},\quad\text{for}\; k\ge1.
\end{equation}
As opposed to the basic quadratic transform and nonhomogeneous quadratic transform that guarantee an error bound of $O(1/k)$, the extrapolated quadratic transform provides an improved error bound of $O(1/k^2)$. Moreover, the extrapolated quadratic transform inherits from the nonhomogeneous quadratic transform the advantage of not requiring large matrix inverse operations. 

\section{Conclusion}
\label{sec:conclusion}

Many problems in signal processing and machine learning naturally lead to
optimization with fractional structure. This article
provides an up-to-date and comprehensive account of an FP technique
termed quadratic transform. As opposed to the classic FP methods (e.g., Dinkelbach's method) that are typically limited to the scalar-valued
single-ratio problem, the quadratic transform allows for a much wider
scope of FP problems, ranging from the sum-of-functions-of-ratio problem to the
matrix-ratio problem. Although the quadratic transform is the main focus,
this article also covers other related FP techniques, including the Lagrangian dual
transform and the AM-GM inequality transform. Moreover, we explore the extensive connections between the quadratic transform and several well-known optimization methods, including the MM method, the gradient projection, the WMMSE algorithm, and the fixed-point iteration method. We further examine the speed of convergence of
the quadratic transform. Aside from the theoretical aspect, this article pays
much attention to a variety of applications of FP, e.g., SVM, unsupervised
data clustering, AoI minimization, power control, beamforming,
and channel estimation.

\section*{Acknowledgments}

The authors wish to thank the anonymous reviewers for many helpful suggestions, especially on the rational optimization \cite{Jibetean} and Lasserre's hierarchy \cite{Lasserre2001}.

The work of Kaiming Shen was supported in part by the National Natural Science Foundation of China (NSFC) under Grant 12426306 and in part by the Guangdong Major Project of Basic and Applied Basic Research under Grant 2023B0303000001. The work of Wei Yu was supported by the Natural Sciences and Engineering Research Council (NSERC) of Canada via a Discovery Grant.

%\bibColoredItems{blue}{multiobj_FP,bilevel_FP,jaberipour2010solving,FettweisTWC2012,Lasserre2001,Dumi2007,Marmin2021,OtterstenSP2019robustClustering,OtterstenTSP2014EE,ZapponeTWC2015,Bugarin2015,shenNIPS,Hien2023,Bertsekas_book,falk1992optimizing,evolutionary}

\bibliographystyle{IEEEtran}
\bibliography{IEEEabrv,strings}

\begin{thebibliography}{10}
\providecommand{\url}[1]{#1}
\csname url@samestyle\endcsname
\providecommand{\newblock}{\relax}
\providecommand{\bibinfo}[2]{#2}
\providecommand{\BIBentrySTDinterwordspacing}{\spaceskip=0pt\relax}
\providecommand{\BIBentryALTinterwordstretchfactor}{4}
\providecommand{\BIBentryALTinterwordspacing}{\spaceskip=\fontdimen2\font plus
\BIBentryALTinterwordstretchfactor\fontdimen3\font minus
  \fontdimen4\font\relax}
\providecommand{\BIBforeignlanguage}[2]{{%
\expandafter\ifx\csname l@#1\endcsname\relax
\typeout{** WARNING: IEEEtran.bst: No hyphenation pattern has been}%
\typeout{** loaded for the language `#1'. Using the pattern for}%
\typeout{** the default language instead.}%
\else
\language=\csname l@#1\endcsname
\fi
#2}}
\providecommand{\BIBdecl}{\relax}
\BIBdecl

\bibitem{charnes_1962}
A.~Charnes and W.~W. Cooper, ``Programming with linear fractional
  functionals,'' \emph{Nav. Res. Logistics Quart.}, vol.~9, no.~3, pp.
  181--186, Dec. 1962.

\bibitem{schaible_transform}
S.~Schaible, ``Parameter-free convex equivalent and dual programs of fractional
  programming problems,'' \emph{Zeitschrift f{\"u}r Oper. Res.}, vol.~18,
  no.~5, pp. 187--196, Oct. 1974.

\bibitem{dinkelbach_transform}
W.~Dinkelbach, ``On nonlinear fractional programming,'' \emph{Manage. Sci.},
  vol. 133, no.~7, pp. 492--498, Mar. 1967.

\bibitem{Jibetean}
D.~Jibetean and E.~de~Klerk, ``Global optimization of rational functions: a
  semidefinite programming approach,'' \emph{Math. Program.}, vol. 106, pp.
  93--109, 2006.

\bibitem{shen2018fractional}
K.~Shen and W.~Yu, ``Fractional programming for communication systems---{Part
  I: Power} control and beamforming,'' \emph{{IEEE} Trans. Signal Process.},
  vol.~66, no.~10, pp. 2616--2630, Mar. 2018.

\bibitem{IM_fp}
I.~M. Stancu-Minasian, \emph{Fractional Programming: {T}heory, Methods and
  Applications}.\hskip 1em plus 0.5em minus 0.4em\relax Kluwer Academic
  Publishers, 1992.

\bibitem{crouzeix1985algorithm}
J.~P. Crouzeix, J.~A. Ferland, and S.~Schaible, ``An algorithm for generalized
  fractional programs,'' \emph{J. Optim. Theory Appl.}, vol.~47, no.~1, pp.
  35--49, Sep. 1985.

\bibitem{freund}
R.~W. Freund and F.~Jarre, ``Solving the sum-of-ratios problem by an
  interior-point method,'' \emph{J. Global Optim.}, vol.~19, no.~1, pp.
  83--102, 2001.

\bibitem{jorswieck}
A.~Zappone and E.~Jorswieck, ``Energy efficiency in wireless networks via
  fractional programming theory,'' \emph{Found. Trends Commun. Inf. Theory},
  vol.~11, no.~3, pp. 185--396, Jun. 2015.

\bibitem{Yannan_TSP24}
Y.~Chen, L.~Zhao, and K.~Shen, ``Mixed max-and-min fractional programming for
  wireless networks,'' \emph{{IEEE} Trans. Signal Process.}, vol.~72, pp.
  337--351, Jan. 2024.

\bibitem{shen2018fractional2}
K.~Shen and W.~Yu, ``Fractional programming for communication systems---{Part
  II}: {U}plink scheduling via matching,'' \emph{{IEEE} Trans. Signal
  Process.}, vol.~66, no.~10, pp. 2631--2644, Mar. 2018.

\bibitem{shen2019optimization}
K.~Shen, W.~Yu, L.~Zhao, and D.~P. Palomar, ``Optimization of {MIMO}
  device-to-device networks via matrix fractional programming: A
  minorization-maximization approach,'' \emph{{IEEE/ACM} Trans. Netw.},
  vol.~27, no.~5, pp. 2164--2177, Oct. 2019.

\bibitem{ZP_MM+}
Z.~Zhang, Z.~Zhao, and K.~Shen, ``Enhancing the efficiency of {WMMSE} and {FP}
  for beamforming by minorization-maximization,'' in \emph{Proc. {IEEE} Int.
  Conf. Acoust., Speech, Signal Process. (ICASSP)}, Jun. 2023.

\bibitem{Shen_JSAC24}
K.~Shen, Z.~Zhao, Y.~Chen, Z.~Zhang, and H.~V. Cheng, ``Accelerating quadratic
  transform and {WMMSE},'' \emph{{IEEE} J. Sel. Areas Commun.}, vol.~42,
  no.~11, pp. 3110--3124, Nov. 2024.

\bibitem{multiobj_FP}
Z.-A. Liang, H.-X. Huang, and P.~M. Pardalos, ``Efficiency conditions and
  duality for a class of multiobjective fractional programming problems,''
  \emph{J. Global Optim.}, vol.~27, pp. 447--471, Dec. 2003.

\bibitem{bilevel_FP}
K.~Mathur and M.~C. Puri, ``On bilevel fractional programming,'' \emph{Optim.},
  vol.~35, no.~3, pp. 215--226, May 1995.

\bibitem{jaberipour2010solving}
M.~Jaberipour and E.~Khorram, ``Solving the sum-of-ratios problems by a harmony
  search algorithm,'' \emph{J. Comput. Appl. Math.}, vol. 234, no.~3, pp.
  733--742, Jun. 2010.

\bibitem{evolutionary}
H.~Arsham and A.~B. Kahn, ``A complete algorithm for linear fractional
  programs,'' \emph{Comput. Math. Appl.}, vol.~20, no.~7, pp. 11--23, 1990.

\bibitem{Bjornson_cellfree21}
S.~Chakraborty, O.~T. Demir, E.~Bj\"{o}rnson, and P.~Giselsson, ``Efficient
  downlink power allocation algorithms for cell-free massive {MIMO} systems,''
  \emph{IEEE Open J. Commun. Soc.}, vol.~2, pp. 168--186, Jan. 2021.

\bibitem{Gu_TWC22_satellite}
P.~Gu, R.~Li, and R.~Tafazolli, ``Dynamic cooperative spectrum sharing in a
  multi-beam {LEO}-{GEO} co-existing satellite system,'' \emph{{IEEE} Trans.
  Wireless Commun.}, vol.~21, no.~2, pp. 1170--1182, Feb. 2022.

\bibitem{larsson_RIS}
H.~Guo, Y.-C. Liang, J.~Chen, and E.~G. Larsson, ``Weighted sum-rate
  maximization for reconfigurable intelligent surface aided wireless
  networks,'' \emph{{IEEE} Trans. Wireless Commun.}, vol.~19, no.~5, pp.
  3064--3076, May 2020.

\bibitem{Zhang_TSP20_mmWave}
J.~Zhang, X.~Hu, and C.~Zhong, ``Phase calibration for intelligent reflecting
  surfaces assisted millimeter wave communications,'' \emph{{IEEE} Trans.
  Signal Process.}, vol.~70, pp. 1026--1040, Feb. 2022.

\bibitem{masouros_isac_twc22}
N.~Su, F.~Liu, Z.~Wei, Y.-F. Liu, and C.~Masouros, ``Secure dual-functional
  radar-communication transmission: {E}xploiting interference for resilience
  against target eavesdropping,'' \emph{{IEEE} Trans. Wireless Commun.},
  vol.~21, no.~9, pp. 7238--7252, Sep. 2022.

\bibitem{Yang_TVT20_ISAC}
J.~Yang, G.~Cui, X.~Yu, and L.~Kong, ``Dual-use signal design for radar and
  communication via ambiguity function sidelobe control,'' \emph{{IEEE} Trans.
  Veh. Technol.}, vol.~69, no.~9, pp. 9781--9793, Sep. 2020.

\bibitem{Palomar_TSP19_shaping}
L.~Wu and D.~P. Palomar, ``Sequence design for spectral shaping via
  minimization of regularized spectral level ratio,'' \emph{{IEEE} Trans.
  Signal Process.}, vol.~67, no.~18, pp. 4683--4695, Sep. 2019.

\bibitem{Lei_TCI24_imaging}
J.~Lei and Q.~Liu, ``Fractional optimization with the learnable prior for
  electrical capacitance tomography,'' \emph{IEEE Trans. Comp. Imag.}, vol.~10,
  pp. 304--317, Feb. 2024.

\bibitem{Beck_SIAM06_QCFQ}
A.~Beck, A.~Ben-Tal, and M.~Teboulle, ``Finding a global optimal solution for a
  quadratically constrained fractional quadratic problem with applications to
  the regularized total least squares,'' \emph{SIAM J. Matrix Anal. Appl.},
  vol.~28, no.~2, pp. 425--445, 2006.

\bibitem{Park_edge_computing_21}
S.-H. Park, S.~Jeong, J.~Na, O.~Simeone, and S.~Shamai, ``Collaborative cloud
  and edge mobile computing in {C-RAN} systems with minimal end-to-end
  latency,'' \emph{IEEE Trans. Signal Inf. Process. Netw.}, vol.~7, pp.
  259--274, Apr. 2021.

\bibitem{guanding_FL}
Y.~He, J.~Ren, G.~Yu, and J.~Yuan, ``Importance-aware data selection and
  resource allocation in federated edge learning system,'' \emph{{IEEE} Trans.
  Veh. Technol.}, vol.~69, no.~11, pp. 13\,593--13\,605, Nov. 2020.

\bibitem{SVM_FP}
B.~Pirouz and M.~Gaudioso, ``New mixed integer fractional programming problem
  and some multi-objective models for sparse optimization,'' \emph{Soft
  Comput.}, vol.~27, pp. 15\,893--15\,904, Jul. 2023.

\bibitem{ZhaoJun_JSAC}
J.~Zhao, L.~Qian, and W.-H. Yu, ``Human-centric resource allocation in the
  metaverse over wireless communications,'' \emph{{IEEE} J. Sel. Areas
  Commun.}, vol.~42, no.~3, pp. 514--537, Mar. 2024.

\bibitem{discriminant_Dias}
S.~Dias, P.~Brito, and P.~Amaral, ``Discriminant analysis of distributional
  data via fractional programming,'' \emph{Eur. J. of Oper. Res.}, vol. 294,
  pp. 206--218, Jan. 2021.

\bibitem{Chen_PAMI}
X.~Chen, Z.~Xiao, F.~Nie, and J.~Z. Huang, ``{FINC: A}n efficient and effective
  optimization method for normalized cut,'' \emph{{IEEE} Trans. Pattern Anal.
  Mach. Intell.}, Feb. 2022.

\bibitem{submodular_Kawahara}
Y.~Kawahara, K.~Nagano, and Y.~Okamoto, ``Submodular fractional programming for
  balanced clustering,'' \emph{Pattern Recognit. Lett.}, vol.~32, pp. 235--243,
  2011.

\bibitem{xiaojun_air}
C.~Zhong, H.~Yang, and X.~Yuan, ``Over-the-air federated multi-task learning
  over {MIMO} multiple access channels,'' \emph{{IEEE} Trans. Wireless
  Commun.}, vol.~22, no.~6, pp. 3853--3868, Jun. 2023.

\bibitem{shen2020enhanced}
K.~Shen, H.~V. Cheng, X.~Chen, Y.~C. Eldar, and W.~Yu, ``Enhanced channel
  estimation in massive {MIMO} via coordinated pilot design,'' \emph{{IEEE}
  Trans. Commun.}, vol.~68, no.~11, pp. 6872--6885, Nov. 2020.

\bibitem{FettweisTWC2012}
C.~Isheden, Z.~Chong, E.~Jorswieck, and G.~Fettweis, ``Framework for link-level
  energy efficiency optimization with informed transmitter,'' \emph{{IEEE}
  Trans. Wireless Commun.}, vol.~11, no.~8, pp. 2946--2957, Aug. 2012.

\bibitem{Bugarin2015}
F.~Bugarin, D.~Henrion, and J.-B. Lasserre, ``Minimizing the sum of many
  rational functions,'' \emph{Math. Program. Comput.}, vol.~8, no.~1, pp.
  83--111, Aug. 2015.

\bibitem{Nie}
J.~Nie, ``Optimality conditions and finite convergence of {Lasserre}'s
  hierarchy,'' \emph{Math. Program.}, vol. 146, pp. 97--121, 2013.

\bibitem{Lasserre2001}
J.-B. Lasserre, ``Global optimization with polynomials and the problem of
  moments,'' \emph{SIAM J. Optim.}, vol.~11, no.~3, pp. 796--817, Jan. 2001.

\bibitem{OtterstenSP2019robustClustering}
A.~Gharanjik, M.~Soltanalian, M.~R.~B. Shankar, and B.~Ottersten,
  ``Grab-n-pull: A max-min fractional quadratic programming framework with
  applications in signal and information processing,'' \emph{Signal Process.},
  vol. 160, Feb. 2019.

\bibitem{kuno}
T.~Kuno, ``A branch-and-bound algorithm for maximizing the sum of several
  linear ratios,'' \emph{J. Global Optim.}, vol.~22, pp. 155--174, 2002.

\bibitem{OtterstenTSP2014EE}
S.~He, Y.~Huang, L.~Yang, and B.~Ottersten, ``Coordinated multicell multiuser
  precoding for maximizing weighted sum energy efficiency,'' \emph{{IEEE}
  Trans. Signal Process.}, vol.~62, no.~3, pp. 741--751, Feb. 2014.

\bibitem{ZapponeTWC2015}
L.~Venturino, A.~Zappone, C.~Risi, and S.~Buzzi, ``Energy-efficient scheduling
  and power allocation in downlink {OFDMA} networks with base station
  coordination,'' \emph{{IEEE} Trans. Wireless Commun.}, vol.~14, no.~1, pp.
  1--14, Jan. 2015.

\bibitem{razaviyayn2013unified}
M.~Razaviyayn, M.~Hong, and Z.-Q. Luo, ``A unified convergence analysis of
  block successive minimization methods for nonsmooth optimization,''
  \emph{SIAM J. Optim.}, vol.~23, no.~2, pp. 1126--1153, 2013.

\bibitem{sun2016majorization}
Y.~Sun, P.~Babu, and D.~P. Palomar, ``Majorization-minimization algorithms in
  signal processing, communications, and machine learning,'' \emph{{IEEE}
  Trans. Signal Process.}, vol.~65, no.~3, pp. 794--816, Aug. 2016.

\bibitem{Yates_JSAC21}
R.~D. Yates, Y.~Sun, D.~R. Brown, S.~K. Kaul, E.~Modiano, and S.~Ulukus, ``Age
  of information: {A}n introduction and survey,'' \emph{{IEEE} J. Sel. Areas
  Commun.}, vol.~39, no.~5, pp. 1183--1210, May 2021.

\bibitem{kaul2018age}
S.~K. Kaul and R.~D. Yates, ``Age of information: Updates with priority,'' in
  \emph{Proc. {IEEE} Int. Symp. Inf. Theory (ISIT)}, Jun. 2018, pp. 2644--2648.

\bibitem{john}
J.~Papandriopoulos and J.~S. Evans, ``S{CALE}: {A} low-complexity distributed
  protocol for spectrum balancing in multiuser {DSL} networks,'' \emph{{IEEE}
  Trans. Inf. Theory}, vol.~55, no.~8, pp. 3711--3724, Aug. 2009.

\bibitem{shenNIPS}
Y.~Chen, B.~Huang, L.~Zhao, and K.~Shen, ``Multidimensional fractional
  programming for normalized cuts,'' in \emph{Conf. Neural Inf. Process. Syst.
  (NeurIPS)}, Dec. 2024, pp. 89\,563--89\,583.

\bibitem{NCut}
J.~Shi and J.~Malik, ``Normalized cuts and image segmentation,'' \emph{{IEEE}
  Trans. Pattern Anal. Mach. Intell.}, vol.~22, no.~8, pp. 888--905, Feb. 2000.

\bibitem{FCD}
F.~Nie, J.~Lu, D.~Wu, R.~Wang, and X.~Li, ``A novel normalized-cut solver with
  nearest neighbor hierarchical initialization,'' \emph{{IEEE} Trans. Pattern
  Anal. Mach. Intell.}, vol.~46, no.~1, pp. 659--666, May 2024.

\bibitem{Nesterov_book}
Y.~Nesterov, \emph{Lectures on Convex Optimization (Second Edition)}.\hskip 1em
  plus 0.5em minus 0.4em\relax Springer, 2018.

\bibitem{Hien2023}
L.~T.~K. Hien, D.~N. Phan, and N.~Gillis, ``An inertial block majorization
  minimization framework for nonsmooth nonconvex optimization,'' \emph{J. Mach.
  Learn. Res.}, vol.~24, no.~18, pp. 1--41, Jan. 2023.

\bibitem{chris}
S.~S. Christensen, R.~Argawal, E.~de~Carvalho, and J.~M. Cioffi, ``Weighted
  sum-rate maximization using weighted {MMSE} for {MIMO-BS} beamforming
  design,'' \emph{{IEEE} Trans. Wireless Commun.}, vol.~7, no.~12, pp. 1--7,
  Dec. 2008.

\bibitem{luo_wmmse}
Q.~Shi, M.~Razaviyayn, Z.-Q. Luo, and C.~He, ``An iteratively weighted {MMSE}
  approach to distributed sum-utility maximization for a {MIMO} interfering
  broadcast channel,'' \emph{{IEEE} Trans. Signal Process.}, vol.~59, no.~9,
  pp. 4331--4340, Sep. 2011.

\bibitem{wang_vt}
L.~Venturino, N.~Prasad, and X.~Wang, ``Coordinated scheduling and power
  allocation in downlink multicell {OFDMA} networks,'' \emph{IEEE Trans. Veh.
  Technol.}, vol.~58, no.~6, pp. 2835--2848, Jul. 2012.

\bibitem{hayssam}
H.~Dahrouj, W.~Yu, and T.~Tang, ``Power spectrum optimization for interference
  mitigation via iterative function evaluation,'' \emph{EURASIP J. Wireless
  Commun. Netw.}, Aug. 2012.

\bibitem{yates}
R.~D. Yates, ``A framework for uplink power control in cellular radio
  systems,'' \emph{{IEEE} J. Sel. Areas Commun.}, vol.~13, no.~7, pp.
  1341--1347, Sep. 1995.

\bibitem{Bertsekas_book}
D.~P. Bertsekas, \emph{Nonlinear Programming (Third Edition)}.\hskip 1em plus
  0.5em minus 0.4em\relax Athena Scientific, 2016.

\end{thebibliography}

\begin{IEEEbiographynophoto}
{Kaiming Shen} received the B.Eng. degree in information security and the B.Sc. degree in mathematics from Shanghai Jiao Tong University, China in 2011, and the Ph.D. degree in electrical and computer engineering from the University of Toronto in 2020. He has been with the School of Science and Engineering at The Chinese University of Hong Kong (Shenzhen), China as a tenure-track assistant professor since 2020. He received the IEEE Signal Processing Society Young Author Best Paper Award in 2021, and the Frontiers of Science Award in 2024. He currently serves as an Editor for IEEE Transactions on Wireless Communications.
\end{IEEEbiographynophoto}

%\vskip 0pt plus -1fil

\begin{IEEEbiographynophoto}
{Wei Yu} received the B.A.Sc. degree in computer engineering and mathematics from the University of Waterloo, Canada, and the Ph.D. degree in electrical engineering from Stanford University, USA. He is currently a Professor in the Electrical and Computer Engineering Department at the University of Toronto, where he holds a Canada Research Chair (Tier 1) in Information Theory and Wireless Communications. He served as the Chair of the Signal Processing for Communications and Networking Technical Committee of the IEEE Signal Processing Society in 2017-2018, and the President of the IEEE Information Theory Society in 2021. Prof. Wei Yu is a Clarivate Highly Cited Researcher. He is a Fellow of IEEE.
\end{IEEEbiographynophoto}

%\vskip 0pt plus -1fil

\end{document}